%% file: main2.tex
\documentclass[english,aps,notitlepage,superscriptaddress,nofootinbib,citeautoscript]{revtex4-1}
\usepackage[T1]{fontenc}
\usepackage[latin9]{inputenc}
\usepackage{color}
\usepackage{babel}
\usepackage{verbatim}
\usepackage{units}
\usepackage{algorithm2e}
\usepackage{amsmath}
\usepackage{graphicx}
\usepackage[unicode=true,pdfusetitle,
 bookmarks=true,bookmarksnumbered=false,bookmarksopen=false,
 breaklinks=false,pdfborder={0 0 0},pdfborderstyle={},backref=false,colorlinks=true]
 {hyperref}
\hypersetup{
 linkcolor=coolblack,citecolor=burntorange,urlcolor=burntgreen}

\makeatletter
\usepackage{color}
\definecolor{burntorange}{rgb}{0.8, 0.33, 0.0}
\definecolor{charcoal}{rgb}{0.21, 0.27, 0.31}
\definecolor{coolblack}{rgb}{0.0, 0.28, 0.49}
\definecolor{burntgreen}{rgb}{0.05, 0.45, 0.27}

\usepackage{lineno}

\usepackage{multibib}

\makeatother

\begin{document}
\include{DCM_new}

\appendix
\setcounter{figure}{0}
\renewcommand{\thefigure}{A\arabic{figure}}
\renewcommand{\theHfigure}{A.\thefigure}

\input{SM.tex}

\end{document}

%% file: DCM_new.tex
\title{From statistical inference to a differential learning rule for stochastic
neural networks}
\author{Luca Saglietti}
\affiliation{Microsoft Research New England, Cambridge (MA), USA}
\affiliation{Italian Institute for Genomic Medicine, Torino, Italy}
\author{Federica Gerace}
\affiliation{Politecnico di Torino, DISAT, Torino, Italy}
\affiliation{Italian Institute for Genomic Medicine, Torino, Italy}
\author{Alessandro Ingrosso}
\affiliation{Center for Theoretical Neuroscience, Columbia University, New York,
USA}
\author{Carlo Baldassi}
\affiliation{Bocconi Institute for Data Science and Analytics, Bocconi University,
Milano, Italy}
\affiliation{Italian Institute for Genomic Medicine, Torino, Italy}
\affiliation{Istituto Nazionale di Fisica Nucleare, Torino, Italy}
\author{Riccardo Zecchina}
\affiliation{Bocconi Institute for Data Science and Analytics, Bocconi University,
Milano, Italy}
\affiliation{Italian Institute for Genomic Medicine, Torino, Italy}
\affiliation{International Centre for Theoretical Physics, Trieste, Italy}
\begin{abstract}
Stochastic neural networks are a prototypical computational device
able to build a probabilistic representation of an ensemble of external
stimuli. Building on the relationship between inference and learning,
we derive a synaptic plasticity rule that relies only on delayed activity
correlations, and that shows a number of remarkable features. Our
\emph{delayed-correlations matching} (DCM) rule satisfies some basic
requirements for biological feasibility: finite and noisy afferent
signals, Dale's principle and asymmetry of synaptic connections, locality
of the weight update computations. Nevertheless, the DCM rule is capable
of storing a large, extensive number of patterns as attractors in
a stochastic recurrent neural network, under general scenarios without
requiring any modification: it can deal with correlated patterns,
a broad range of architectures (with or without hidden neuronal states),
one-shot learning with the palimpsest property, all the while avoiding
the proliferation of spurious attractors. When hidden units are present,
our learning rule can be employed to construct Boltzman-Machine-like
generative models, exploiting the addition of hidden neurons in feature
extraction and classification tasks. 
\end{abstract}
\maketitle

\tableofcontents{}

\section*{Introduction}

One of the main open problems of neuroscience is understanding the
learning principles which enable our brain to store and process information.
Neural computation takes place in an extremely noisy environment:
experiments show that various sources of variability and fluctuations
make neurons, synapses, and neural systems intrinsically stochastic
\citep{RollsDecoNoisy}. Such internal noise can originate at different
levels, for instance from the unreliable transmission of synaptic
vesicles, from the random opening and closing of ion channels or from
the trial-to-trial variability in neural responses to external stimuli
\citep{cannon2010stochastic,flight2008synaptic,azouz1999cellular,gerstner2002spiking,brascamp2006time}.
At the same time, even the typical sensory input is often blurry and
ambiguous. A probabilistic inference framework is thus the natural
choice for modeling all the uncertainties affecting neural learning
\citep{buesing2011neural}.

A widespread belief is that learning occurs at the synaptic level,
both in terms of creation of new connections and by synaptic strength
potentiation or depression \citep{bliss1993synaptic,rogan1997fear,whitlock2006learning}.
Synaptic plasticity can be encoded in a learning principle that relates
the modulation of the efficacy of a synapse to its pre- and post-synaptic
neural activity. The simplest synaptic plasticity rule, Hebb's rule,
states that positive correlation between pre- and post-synaptic spikes
leads to long-term potentiation (LTP), while negative correlation
induces long-term depression (LTD). One important feature of Hebbian
plasticity is its capability to shape the connectivity of neural architectures
in a way that captures the statistics of the stimuli. This issue has
been addressed in a number of modeling studies, starting from the
classical theory of development of neural selectivity \citep{Ben-Yishai1995},
to more modern accounts of neural tuning that use homeostasis-stabilized
Hebbian plasticity in large spiking network models \citep{ClopathEmergence}. 

On the other hand, it has long been recognized that Hebbian plasticity
is capable of generating attractor dynamics in a variety of recurrent
architectures: the concept of attractor neural network is one of the
most important in modern neuroscience, in that it can account for
a variety of neurophysiological observations of persistent activity
in various brain areas. Examples include line attractor (neural integrator)
models in oculomotor control \citep{compte2000synaptic}, ring attractor
models in head direction systems \citep{zhang1996representation},
and a plethora of models of persistent neural activity whose common
feature is a local connectivity pattern which stabilizes bump attractors
by means of lateral inhibition. 

The main intuition that led to the introduction of the prototypical
model of attractor network -- the Hopfield model -- was that the
\emph{frustration phenomenon} in disordered systems (spin glasses),
namely the proliferation of metastable states due to the strongly
heterogeneous nature of the couplings, could be exploited to embed
uncorrelated patterns as steady states of a network dynamics. In the
Hopfield model, a straightforward application of Hebb's rule leads
to a definition for the synaptic weights that allows for an extensive
number of attractors to be stored, but exhibits a phenomenon known
as \emph{catastrophic forgetting }\citep{Amit:1989:MBF:77051}: all
memories are lost, due to the existence of an absorbing spin glass
state uncorrelated with the memories, as soon as the maximum number
of attractors is exceeded. Since the original introduction of the
Hopfield model \citep{Hopfield1982}, many generalized Hebb rules
have been proposed, able to deal with sparse patterns or low activity
levels, see e.g.~ref.~\citep{Amit:1989:MBF:77051}. Moreover, Hebbian
learning has been profitably used to embed attractor states in a variety
of neural network models spanning from binary units to graded neurons
(rate models) \citep{Hopfield1984} and spiking networks \citep{amit1997model}.

Different lines of research concerning attractor neural networks in
Statistical Mechanics and Computational Neuroscience have strong ties
with the study of generative energy-based models: the formalism of
Boltzmann Machines allows for a generalization to neural networks
with hidden neural states \citep{hinton1983_BM,hinton1985_BMlearning}.
This introduction, though, comes at the price of serious technical
complications in the definition of a viable learning rule. Some of
these models have become popular also in the machine-learning community,
after proving themselves as useful tools in several deep-learning
applications. This stimulated the development of various learning
heuristics, the most renown being Contrastive Divergence (CD) \citep{carreira2005contrastive_RBM},
and inference methods \citep{gabrie2015training_TAPBM,tanaka1998meanfield_BM,yasuda2012learning_PSL_Completed,kappen1998efficient_MF_2}. 

An alternative direction of research is motivated by many inference
problems in biological systems, where couplings are typically asymmetric
and possibly time-varying. The study of the dynamics and learning
in these purely kinetic models is complicated by the lack of analytical
control over the stationary distribution \citep{KappenMeanField,RoudiHertzMeanField,mezard2011exact}
: a number of interesting mean field techniques based on generalization
of TAP equations have been proposed \citep{KappenMeanField,RoudiHertzMeanField,mezard2011exact}
in this context.

In the present study, we approach many of these problems from a unified
perspective: the main goal of the paper is to devise a biologically
plausible learning rule which could allow a general stochastic neural
network to construct an internal representation of the statistical
ensemble of the stimuli it receives. In the following, we consider
the case of asymmetric synaptic couplings and derive a learning scheme
in which the updates involve only purely local and possibly noise-affected
information. The proposed plasticity rule does not rely on the presence
of supervisory signals or strong external stimuli, and proves to be
compatible with Dale's principle, which requires the homogeneity of
the neurotransmitters released by one neuron across its synaptic terminals
\citep{strata1999dale,catsigeras2013dale}. In this work, we define
the learning process in an on-line context, and our analysis will
be restricted to the case of discrete time dynamics.

In the \textbf{Results} section, for clarity of exposition we will
mostly focus on the specific case of fully-visible neural networks,
giving only a brief overview on the extension to networks comprising
hidden neurons. This last setting is largely expanded upon in the
\textbf{}\textbf{Appendix}, where we also provide further analytical
insights and the implementation details of our numerical experiments.

\section*{Results}

We present our main results in three different subsections. In the
first one (\textbf{The model}), we derive the new plasticity rule
in a framework that encompasses a wide variety of unsupervised and
semi-supervised problems, such as the construction of attractor networks
and learning in generative models with more complicated structures.
In the second one (\textbf{Fully visible case}), we specialize to
the case of attractor networks containing only visible neurons. After
describing a link with the maximum pseudo-likelihood method, we study
the numerical performance of the new learning rule in various settings,
showing how it deals with finite external fields, different coding
levels and the constraint of Dales's law. We then test the rule in
the case of correlated memories, we investigate its proneness to create
spurious attractors and we measure its palimpsest capacity. Finally,
in the third section (\textbf{Adding neuronal states}) we give an
introduction to the more complex case of stochastic networks with
hidden neurons, and review some of the results, presented in \textbf{Appendix~\ref{sec:Visible-to-hidden}},
that were obtained in this setting. 

\subsection*{The model}

We consider the customary simple setup of a network of $N$ stochastic
binary neurons $s=\left\{ s_{i}\right\} _{i=1}^{N}$, with each $s_{i}$
either in $\left\{ -1,+1\right\} $ or $\left\{ 0,1\right\} $, connected
by a set of asymmetric synaptic weights $J_{ij}\neq J_{ji}$, which
evolves with a discrete-time synchronous dynamics described by the
Glauber transition probability: the next state $s^{\prime}$ of the
system depends on the current state $s$ according to the following
factorized probability distribution
\begin{equation}
P\left(s^{\prime}\mid s;\beta\right)=\prod_{i=1}^{N}\sigma\left(s_{i}^{\prime}\mid h_{i};\beta\right),\label{eq:Glauber_dynamics}
\end{equation}
with $\sigma\left(\cdot\right)$ being a sigmoid-shaped neural activation
function defined by $\sigma\left(s\mid h;\beta\right)\propto e^{-\beta sh}$
(the proportionality constant being set by normalization), and $h_{i}=h_{i}^{\mathrm{ext}}+\sum_{j\neq i}J_{ij}s_{j}-\theta_{i}$
being the total neural current -- or \emph{local field} -- obtained
by adding up the recurrent contributions from other neurons to the
external stimulus $h_{i}^{\mathrm{ext}}$. The quantity $\theta_{i}$
serves as a local threshold. The dynamics of the system is thus stochastic,
and the parameter $\beta$ (which has the role of an inverse temperature
in analogous physical models) provides a measure of the dynamical
noise in the system. When the synaptic couplings $J$ are finite and
the external fields are time independent, the dynamics is known to
be ergodic and a steady state defined by a unique stationary distribution
is approached \citep{KappenMeanField}. However, the analytical form
of this steady state distribution is not known for general asymmetric
kinetic models of the type we consider here.

In the following, we formulate the problem of learning as an unsupervised
task where the network has to adapt its parameters in accordance with
some plasticity rule: the goal is to learn an internal representation
of a target probability distribution, which is to be inferred from
a set of external stimuli conveyed to a subset of the neurons. Suppose
we are given a time-independent binary pattern, a vector $\xi$ of
length $N_{\mathcal{V}}\le N$ with components $\xi_{i}=\pm1$, to
be learned by the neural network. This pattern is presented to a group
$\mathcal{V}=\left\{ 1,\dots,N_{\mathcal{V}}\right\} $ of ``visible''
neurons in the form of an external field of variable intensity $\lambda^{\mathrm{ext}}$
in the direction of $\xi$, i.e.~$h_{i}^{\mathrm{ext}}=\lambda^{\mathrm{ext}}\xi_{i}$
for $i\in\mathcal{V}$, while the complementary subset $\mathcal{H}=\left\{ N_{\mathcal{V}}+1,\ldots,N\right\} $
of ``hidden'' neurons receive no external input. We want to model
the scenario in which the stimulus intensity is high (although not
as large as to clamp the neurons) at the onset, and rapidly vanishes.
The initial presence of the field biases the dynamics of the system;
in the retrieval phase, if the stimulus $\xi$ is sufficiently close
to a pattern $\xi^{\prime}$ that the network has learned, the stationary
probability distribution of the visible neuronal states should get
focused in the direction of $\xi^{\prime}$ even after the stimulus
is no longer present.

For the sake of comparison, the classical Hopfield network with Hebbian
learning can be framed in the same setting, as follows: we assume
that there are no hidden neurons, and the dynamics of the stimulus
presentation is a simple two-step process in which the stimulus intensity
$\lambda^{\mathrm{ext}}$ is initially effectively infinite (such
that the other components of the inputs become irrelevant and the
dynamics of the neurons becomes deterministic and fixed, i.e.~such
as to clamp the network) and then drops to $0$. The learning rule
in that case actually only uses the information about the state of
the network during the clamped phase: $\Delta J_{ij}\propto s_{i}s_{j}$
where $s=\xi$ as a consequence of the clamping. In the retrieval
tests, the clamped phase is used to initialize the network, which
subsequently evolves by its own internal dynamics in absence of further
stimuli.

In our framework, we exploit the dynamics of the stimulus during the
learning phase, extracting the correlations that the stimulus induces
on the network dynamics and using them to train the network: since
the final goal of the network is to learn from the driving effect
of the external field, we may require the dynamical evolution in the
freely evolving network to maximally resemble the stimulus-induced
evolution. Intuitively, this amounts at training the network to compensate
the gradual vanishing of the external field by adapting its own recurrent
connections. This requirement can be framed formally as the minimization
of a Kullback-Liebler (KL) divergence between two different conditional
probability distributions corresponding to different levels of intensity
of the external field, $P\left(s^{\prime}|s;\lambda_{1}^{\mathrm{ext}}\right)$
and $P\left(s^{\prime}|s;\lambda_{2}^{\mathrm{ext}}\right)$, averaged
over some initial state probability distribution $P\left(s\right)$.
The analytical details can be found in \textbf{\textbf{Appendix~\ref{sec:Analytic-derivation}}}.
As explained in more detail below, the distribution $P\left(s\right)$
is supposed to be concentrated (for the visible part of the network)
around the direction of the pattern $\xi$, such that $s^{\prime}$
will also be concentrated around $\xi$ as the combined effect of
the initial conditions, the external field and of the recurrent connections;
when the effect of the external field decreases, the recurrent connections
will tend to compensate for this. If these conditions can be met,
then the procedure can be applied repeatedly.

As an initial simplified case, consider the same setting as the Hebbian
learning, i.e. the limiting case of an infinite $\lambda^{\mathrm{ext}}$,
in which the visible part of the network dynamics is clamped. The
\emph{stationary} probability distribution can thus be factorized
over the visible neuronal states $s_{\text{\ensuremath{\mathcal{V}}}}$:
\begin{eqnarray}
P_{\textrm{clamp}}\left(s;\xi\right) & = & \left(\prod_{i\in\mathcal{V}}\delta_{s_{i},\xi_{i}}\right)\,P\left(s_{\mathcal{H}}|s_{\text{\ensuremath{\mathcal{V}}}}\right),\label{eq:clamped}
\end{eqnarray}
where $\delta_{x,y}$ denotes the Kronecker delta symbol which equals
1 if $x=y$ and $0$ otherwise. Here, the conditional probability
of the hidden neuronal states $s_{\mathcal{H}}$, given the visible,
cannot be written explicitly without losing generality. In our learning
scheme, we seek to minimize the difference between the initial (fully
clamped) situation and the subsequent zero-field situation; this requirement
produces the following simple learning rules for the synaptic couplings
and the thresholds:

\begin{eqnarray}
\Delta J_{ij} & \propto & \left(\left\langle s_{i}^{\prime}s_{j}\right\rangle _{\textrm{clamp},\infty}-\left\langle s_{i}^{\prime}s_{j}\right\rangle _{\textrm{clamp},0}\right)\label{eq:learnparameters}\\
\Delta\theta_{i} & \propto & -\left(\left\langle s_{i}^{\prime}\right\rangle _{\textrm{clamp},\infty}-\left\langle s_{i}^{\prime}\right\rangle _{\textrm{clamp},0}\right),\nonumber 
\end{eqnarray}
where $s$ and $s^{\prime}$ denote two successive states of the network,
as above, and $\left\langle \cdot\right\rangle _{\textrm{clamp},\lambda^{\mathrm{ext}}}$
is defined as an average over the possible dynamical responses starting
from a state sampled from $P_{\textrm{clamp}}$:

\begin{eqnarray}
\left\langle s_{i}^{\prime}s_{j}\right\rangle _{\textrm{clamp},\lambda^{\mathrm{ext}}} & = & \sum_{s^{\prime},s}s_{i}^{\prime}s_{j}\,P\left(s_{i}^{\prime}\mid s;\lambda^{\mathrm{ext}}\right)P_{\textrm{clamp}}\left(s;\xi\right)\\
\left\langle s_{i}^{\prime}\right\rangle _{\textrm{clamp},\lambda^{\mathrm{ext}}} & = & \sum_{s^{\prime},s}s_{i}^{\prime}\,P\left(s_{i}^{\prime}\mid s;\lambda^{\mathrm{ext}}\right)P_{\textrm{clamp}}\left(s;\xi\right)\nonumber 
\end{eqnarray}
In the limiting case we simply have $\left\langle s_{i}^{\prime}s_{j}\right\rangle _{\textrm{clamp},\infty}=\xi_{i}\xi_{j}$
for the visible neurons. In general, however, efficiently obtaining
an accurate estimate of this average can pose serious technical challenges.

\begin{figure}
\centering{}\includegraphics[width=0.9\textwidth]{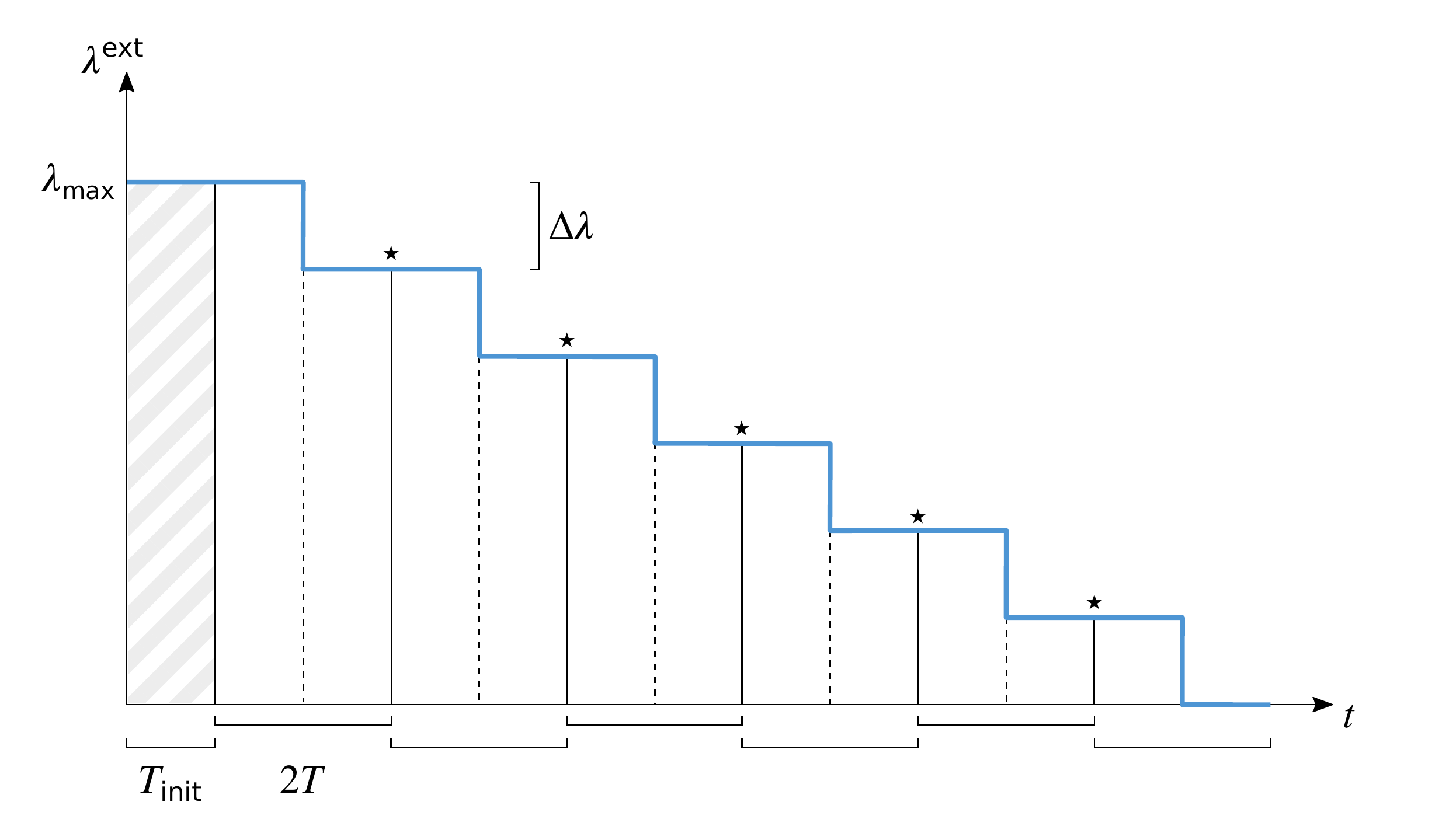}\caption{\label{fig:protocol}\emph{DCM Learning protocol scheme.} This represents
the learning process for one pattern presentation. The blue curve
shows the stepwise dynamics of the external field $\lambda^{\mathrm{ext}}$
as a function of the time $t$ of the network dynamics. The first
time period ($T_{\mathrm{init}}$ time steps, shaded) serves for initializing
the network state in the proximity of the pattern. The protocol then
proceeds in windows of $2T$ steps, each one divided in two phases.
In the middle of each window the field intensity drops by $\Delta\lambda$.
The time-delayed correlations are recorded separately for the two
phases. The parameters are updated at the end of each window, in correspondence
of the $\star$ symbols, according to eq.~(\ref{eq:bio_formulation}).}
\end{figure}
Since the case of a clamping stimulus is biologically unrealistic,
we explore a setting in which the amplitude of the external signal
is comparable to the recurrent contribution exerted by the surrounding
neurons: instead of trying to match the dynamical response of a clamped
model with a freely-evolving one, we introduce a learning protocol
based on a time dependent field intensity $\lambda^{\mathrm{ext}}\left(t\right)$,
which decreases to zero starting from a finite initial value $\lambda_{\max}$.
In the following we will consider a staircase signal intensity $\lambda^{\mathrm{ext}}\left(t\right)$,
lowered by a fixed amount $\Delta\lambda$ after every $2T$ steps
of the time-discretized network dynamics (see fig.~\ref{fig:protocol}).
We should remark, however, that the results presented hereafter are
quite robust with respect to variations in the precise details of
the dynamical protocol for the field, and that the above choice was
purely made for simplicity of presentation and analysis.

The training protocol prescribes the network to try and match its
dynamical behavior at a given level of the field $\lambda$ with that
at a lower level $\lambda-\Delta\lambda$, where the dynamical behavior
is measured in terms of the time-delayed correlations between neurons
\begin{eqnarray}
\left\langle s_{i}^{\prime}s_{j}\right\rangle _{\textrm{\ensuremath{\lambda}}} & = & \sum_{s^{\prime},s}s_{i}^{\prime}s_{j}\,P\left(s_{i}^{\prime}|s;\lambda\right)P\left(s\right)
\end{eqnarray}
and $P\left(s\right)$ is some initial probability distribution roughly
concentrated around the presented pattern $\xi$ for the visible neurons.
More precisely, we suppose that the overall distribution $P\left(s_{i}^{\prime}|s;\lambda\right)P\left(s\right)$
induces a dynamics which is confined around $\xi$ and ergodic within
such region. When that is the case, sampling the temporal averages
as the system evolves can provide an estimate of the averages involved
in the above expression. It is reasonable to assume, and confirmed
by our experiments, that this condition will be satisfied if the initial
field $\lambda_{\max}$ is sufficiently large, thus creating an effective
basin of attraction, and if the system evolution manages to keep this
confinement in place even when the field is decreased by adapting
the recurrent connections.

Our learning protocol is thus defined as follows (see fig.~\ref{fig:protocol}):
the network will first record for $T$ time steps its time-delayed
correlations at a given value of the field $\lambda^{\mathrm{ext}}\left(t\right)=\lambda$;
then, it will do the same for another $T$ steps at a lower level,
$\lambda^{\mathrm{ext}}\left(t+T\right)=\lambda-\Delta\lambda$, after
which it will adjust its parameters such as to try to match the two
sets of measurements (see below). The protocol will then restart with
the same field $\lambda^{\mathrm{ext}}\left(t+2T\right)=\lambda-\Delta\lambda$
(but with updated network parameters), proceeding in this way until
the field has dropped to zero. The network state is never reset during
these steps; rather, it keeps following the dynamics of eq.~(\ref{eq:Glauber_dynamics}).
An extra initial period of $T_{\mathrm{init}}$ steps (we generally
set $T_{\mathrm{init}}=T$ in our simulations) at $\lambda_{\max}$
field is used to prepare the network and bias it in the direction
of the pattern.

Therefore, in this approximation, we obtain a new plasticity rule
(the notation $\left\langle \cdot\right\rangle _{t,\lambda}$ here
denotes empirical averages over time in presence of a given field
$\lambda$, and we switch to using $t$ and $t+1$ to denote two consecutive
time steps): 

\begin{eqnarray}
\Delta J_{ij} & \propto & \left(\left\langle s_{i}^{t+1}s_{j}^{t}\right\rangle _{t,\lambda}-\left\langle s_{i}^{t+1}s_{j}^{t}\right\rangle _{t,\lambda-\Delta\lambda}\right)\label{eq:bio_formulation}\\
\Delta\theta_{i} & \propto & -\left(\left\langle s_{i}^{t+1}\right\rangle _{t,\lambda}-\left\langle s_{i}^{t+1}\right\rangle _{t,\lambda-\Delta\lambda}\right),\nonumber 
\end{eqnarray}
which simply tries to match the time-delayed correlations in the consecutive
time windows, until the signal has vanished and the system evolves
freely. All the needed information is thus local with respect to each
synapse. In order to learn a given extensive set of $\alpha N$ patterns,
the same procedure has to be repeated cyclically: a pattern is presented
with decreasing intensity while the network adapts its parameters,
then the network moves to the next pattern. The network is not reset
even between one pattern and the next. We call this learning rule
``delayed-correlations matching'', DCM for short. The full algorithm
is detailed in \textbf{\textbf{Appendix~\ref{sec:Simulation}}} together
with the corresponding pseudo-code. 

It is not necessary for the field dynamics to end up exactly at zero
intensity: following the same idea proposed in ref. \citep{baldassi2018inverse},
the learning scheme described above can be made more robust if one
requires the network to face the presence of an antagonist field,
that tries to interfere with the drawing effect of the basin of attraction.
By considering a negative minimal intensity $\lambda_{\text{min}}<0$,
one can in fact both speed up the learning process and induce larger
basins of attraction. If instead the aim is to learn new basins of
attraction coherently, trying not to affect the previously stored
memories, it can be useful to choose a positive $\lambda_{\text{min}}>0$:
this ensures that the sampling process doesn't leave the neighborhood
of the presented pattern, risking to end up in a different memory
and possibly delete it (we will consider this prescription in the
\emph{one-shot learning} scenario).

\subsection*{Fully visible case}

When a network with no hidden neurons is considered ($N_{\mathcal{V}}=N$),
the learning problem effectively reduces to that of constructing a
stochastic attractor neural network with binary units. Kinetically
persistent neuronal states can be indeed observed even with asymmetric
synaptic couplings $J$. We will require the network to embed as stable
and attractive memories an extensive set of i.i.d. random binary $\pm1$
patterns, denoted by $\left\{ \xi^{\mu}\right\} _{\mu=1}^{M}$, with
$M=\alpha N$ (each $\xi^{\mu}$ is an $N$-dimensional vector, and
$\mu$ represents a pattern index). The number of stored patterns
per neuron $\alpha$ is the so called \emph{storage load} of the network.

Since the learning procedure is defined as a cyclical minimization
of a KL-divergence evaluated at the $M$ patterns, the limiting case
with just two dynamical steps and infinite initial field considered
in eq.~(\ref{eq:learnparameters}) can here be reinterpreted exactly
as an on-line optimization of the so called log-pseudo-likelihood: 

\begin{equation}
\mathcal{L}\left(\left\{ \xi^{\mu}\right\} |J_{ij},\theta;\beta\right)=\frac{1}{M}\sum_{\mu=1}^{M}\sum_{i=1}^{N}\log P\left(s_{i}=\xi_{i}^{\mu}|\left\{ s_{j}=\xi_{j}^{\mu}\right\} _{j\neq i};\lambda^{ext}=0\right),
\end{equation}
which is most frequently found in an inference framework \citep{nguyen2017inverse,aurell2012inverse_PSL},
where the parameters of a generative model have to be inferred from
a finite set of complete observations (see \textbf{\textbf{Appendix~\ref{subsec:pseudo-likelihood}}}).

In this case, the update for the synaptic couplings can be written
more explicitly and allows for a clear comparison with the standard
Hebbian plasticity rule:

\begin{eqnarray}
\Delta J_{ij} & \propto & \left(\xi_{i}^{\mu}\xi_{j}^{\mu}-\sum_{s_{i}^{t+1}}P\left(s_{i}^{t+1}|\left\{ s_{j}=\xi_{j}^{\mu}\right\} _{j\neq i};\lambda^{ext}=0\right)s_{i}^{t+1}\xi_{j}^{\mu}\right).\label{eq:update_rule}
\end{eqnarray}

The DCM rule is explicitly asymmetric, and its differential form produces
a homeostatic mechanism constantly trying to reproduce externally
induced correlations in the network dynamics. While in the initial
stages of the learning process the synaptic weights are modified according
to a typical Hebbian prescription - potentiation in case of positive
correlations and depression with negative ones - the comparator effectively
avoids the possibly uncontrolled positive feedback loop of the Hebbian
principle: no change in synapses will occur when the correlations
in the absence of the stimulus already match the ones of the learned
patterns. Incidentally, we also note that in the noise-free limit
$\beta\to\infty$ the perceptron learning rule is recovered (see \textbf{Appendix~\ref{subsec:perceptron}}).
In the case of $s_{i}\in\left\{ -1,+1\right\} $ neurons, we studied
numerically the trend of the maximum storage load achievable with
the DCM rule as a function of the required width of the basins of
attraction. We introduced an operative measure of the basin size,
relating it to the level of corruption of the memories before the
retrieval: a set of $M=\alpha N$ patterns is considered to be successfully
stored at a noise level $\chi$ if, initializing the dynamics in a
state where a fraction $\chi$ of the pattern is randomly corrupted,
the retrieval rate for each pattern is at least $90\%$ (for additional
details see Appendix~\ref{subsec:Basins-width}). In fig.~\ref{fig:critical_capacity}
we compare the DCM rule with the Hopfield model, which is known to
achieve a maximum storage load of $\sim0.14N$.

\begin{figure}
\centering{}\includegraphics[width=0.9\textwidth]{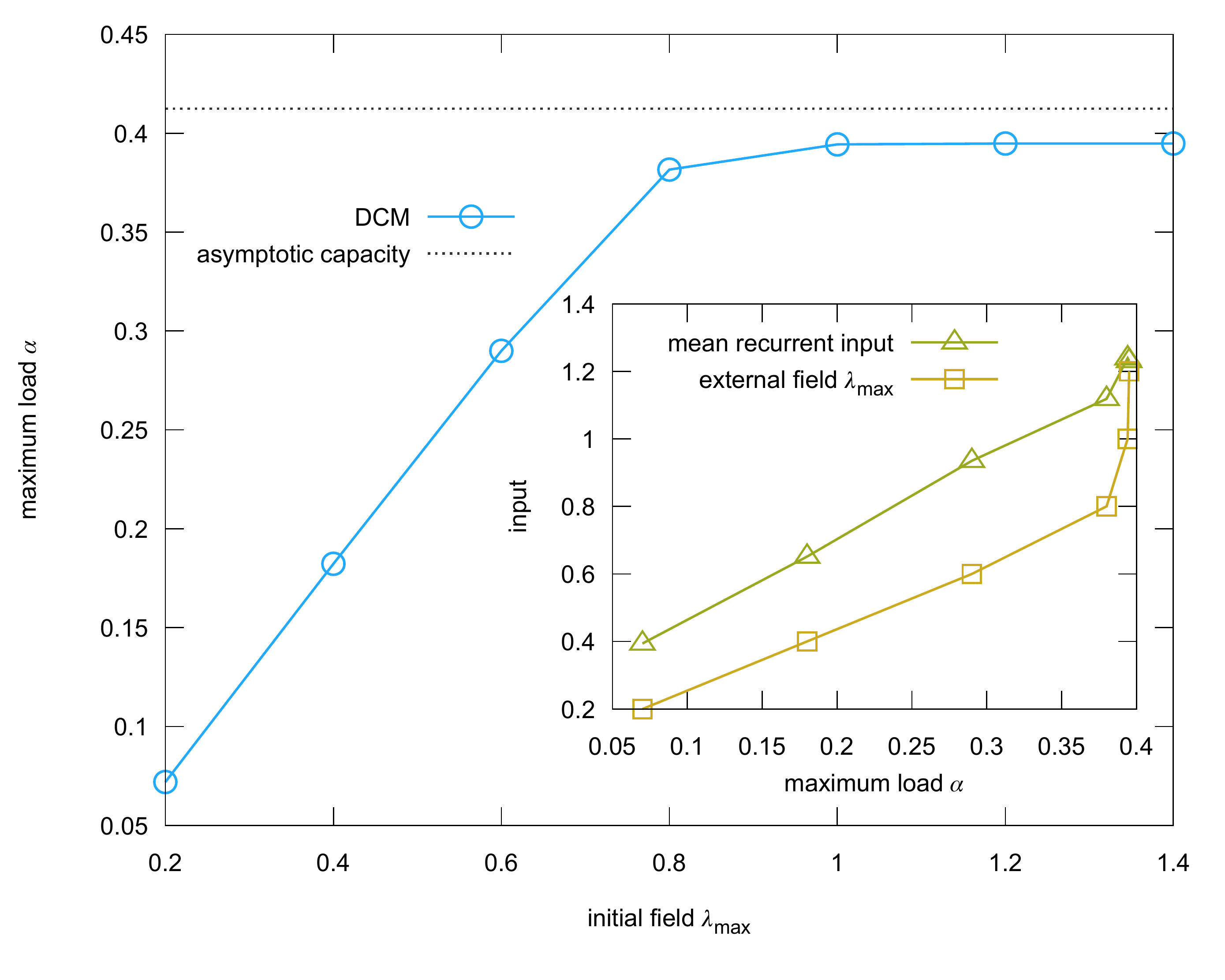}\caption{\emph{\label{fig:capacity_fields}Maximum storage load as a function
of the field intensity} for a network of $N=400$ neurons. The correlations
were recorded in windows of $T=20$ time steps and the field intensity
step was $\Delta\lambda=\lambda_{max}/3$. The noise level in the
retrieval phase is set to $\chi=0.3$ and the temperature to $\beta=2$.
The curve was obtained by averaging over $100$ samples. The inset
shows a comparison between the recurrent and external components of
the inputs, for the same data points of the main panel. The mean recurrent
input was computed as the square root of the mean values. This shows
that the DCM rule is effective even for relatively small stimuli.}
\end{figure}

If we move to the more biologically plausible scenario of finite time-dependent
external fields (eq.~(\ref{eq:bio_formulation})), we clearly see
in fig.~\ref{fig:capacity_fields} that an infinite signal is actually
redundant. If the external field intensity is high enough, the recorded
time-delayed correlations carry enough information about the pattern
to be learned. If instead the signal component in the local field
is dominated by the recurrent contribution from other neurons the
dynamics becomes completely noisy. Since the average strength of the
connections between the neurons increases with the number of stored
memories, the maximum storage load grows with the signal amplitude.
Nevertheless, the results of pseudo-likelihood are already almost
saturated at small fields intensities $\lambda_{\max}\sim1$, and
the DCM rule generally works well even when the stimulus intensity
is relatively small compared to the total recurrent input (see inset
of fig.~\ref{fig:capacity_fields}). The implementation details are
described in \textbf{\textbf{Appendix~\ref{sec:Simulation}}}.

\begin{figure}
\centering{}\includegraphics[width=0.9\textwidth]{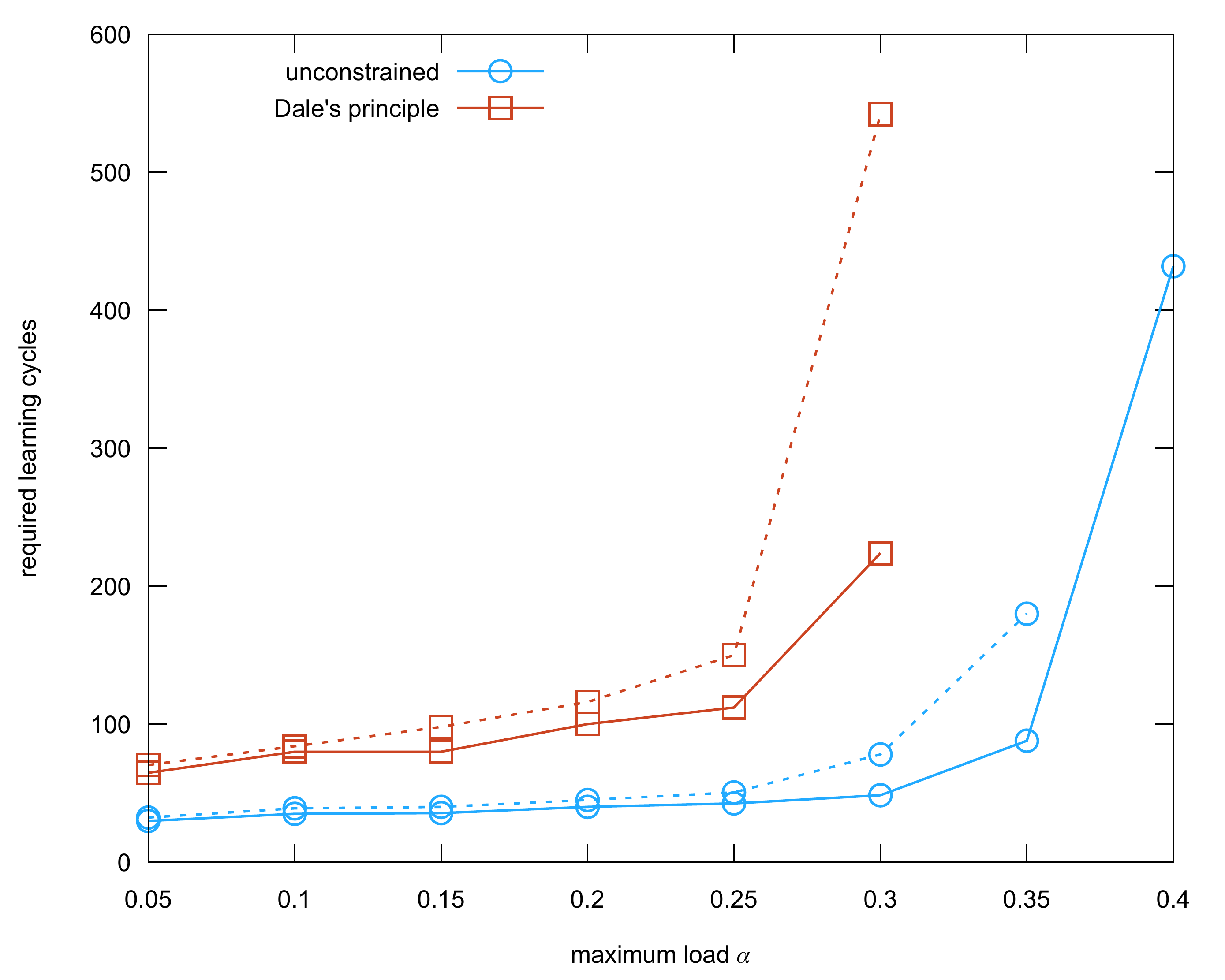}\caption{\emph{\label{fig:learning_cycles}Required learning cycles as a function
of the storage load, for unconstrained and constrained synapses }for
networks of size $N=200$ (dashed curves) and $N=400$ (full curves).
The results for the case of unconstrained synapses (blue curves) and
that of synapses satisfying Dale's principle (red curves) are compared.
Here the chosen inhibitory scheme is the soft ``winner takes all''
mechanism. The noise level in the retrieval phase was set to $\chi=0.3$,
while the sparsity was fixed at $f_{v}=0.5$ in order to avoid finite
size effects with the relatively small networks. The curves are interrupted
at the value of $\alpha$ where the algorithm starts failing.}
\end{figure}

We also considered an alternative model with somewhat more biologically
plausible features, using $s_{i}\in\left\{ 0,1\right\} $ neurons
(see \textbf{\textbf{Appendix~\ref{sec:Types-of-neurons}}}) and
sparse $\xi_{i}^{\mu}\in\left\{ 0,1\right\} $ patterns, and forcing
the synapses to satisfy Dale's law. This means that two sub-populations
of excitatory and inhibitory neurons should be defined, the sign of
their outgoing synapses being fixed a priori. Note that this restriction
reduces the theoretical maximum capacity of the network, although
not dramatically (roughly by half \citep{alemithreshold}). For simplicity,
we restricted our analysis to the case where only excitatory synapses
are plastic and a separate inhibitory sub-network provides a feedback
regulatory effect, whose goal is to maintain the network activity
$S^{t}=\sum_{i=1}^{N}s_{i}^{t}$ around a desired level $Nf_{v}$
(the same sparsity level as the learned patterns), and preventing
epileptic (all-on) or completely switched off states. We tested three
different effective models that implement an inhibitory feedback mechanism:
\emph{(i)} a generalization of the global inhibitory mechanism described
in ref.~\citep{alemithreshold}, tuned such as to counterbalance
the oscillations of the network activity around the desired level;
\emph{(ii)} a soft ``winner-takes-all'' mechanism, effectively playing
the role of a global inhibitory unit \citep{binas2014learning,douglas1989canonical,mountcastle1997columnar,binzegger2004quantitative,douglas2007recurrent,carandini2012normalization,handrich2009biologically,mao2007dynamics,lynch2016computational,oster2006spiking,fang1996dynamics};
\emph{(iii)} a model with adaptive thresholds that allow the Dale's
principle model to behave approximately like the unconstrained one.
The details, including the derivation of the parameters of all these
schemes, are reported in \textbf{\textbf{Appendix~\ref{sec:Inhibitory-Network-models}}}.
For all of them, the results are comparable to the ones shown in fig.~\ref{fig:learning_cycles}.

\begin{figure}[h]
\begin{centering}
\includegraphics[width=0.9\textwidth]{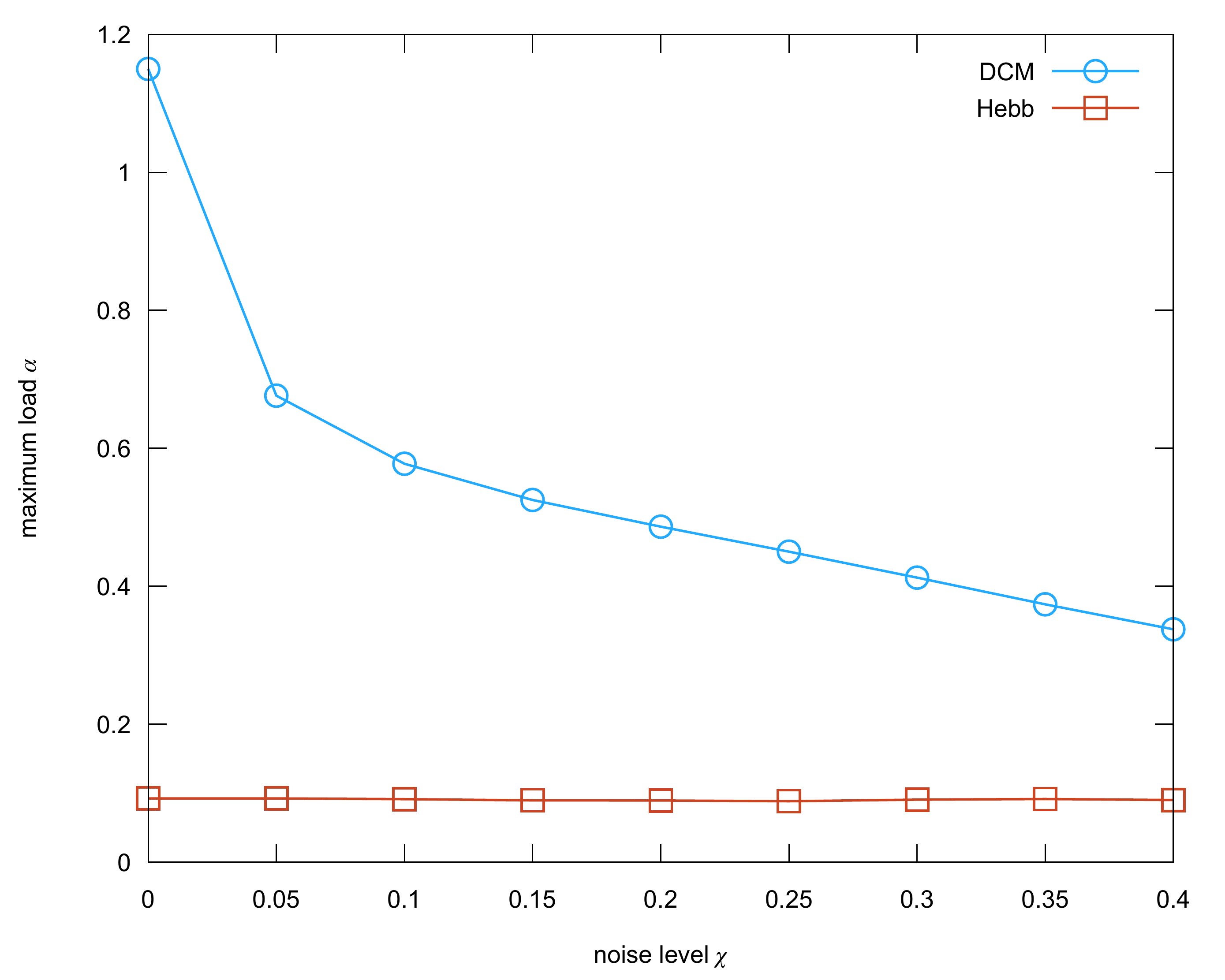}
\par\end{centering}
\centering{}\caption{\emph{\label{fig:critical_capacity}Maximum storage load as a function
of the width of the basin of attraction} for a network of $N=400$
visible neurons. The red and blue curves show the results for Hopfield
model and the DCM rule, respectively. The noise level $\chi$ operatively
measures the width of the basins of attraction (it is the fraction
of corrupted bits that the network is able to correct, see the text
for a more precise definition). Each curve is an average over $10$
samples (error bars are smaller than point size). The inverse temperature
parameter is set to $\beta=2$ in order to fall within the retrieval
phase of the Hopfield model. The critical capacity at zero temperature
is lower than the Gardner bound, $\alpha_{c}=2$, because of the stochastic
component of the dynamics. }
\end{figure}

\subsubsection*{Comparison with Hebbian plasticity rule}

\begin{figure}[h]
\centering{}\includegraphics[width=0.9\textwidth]{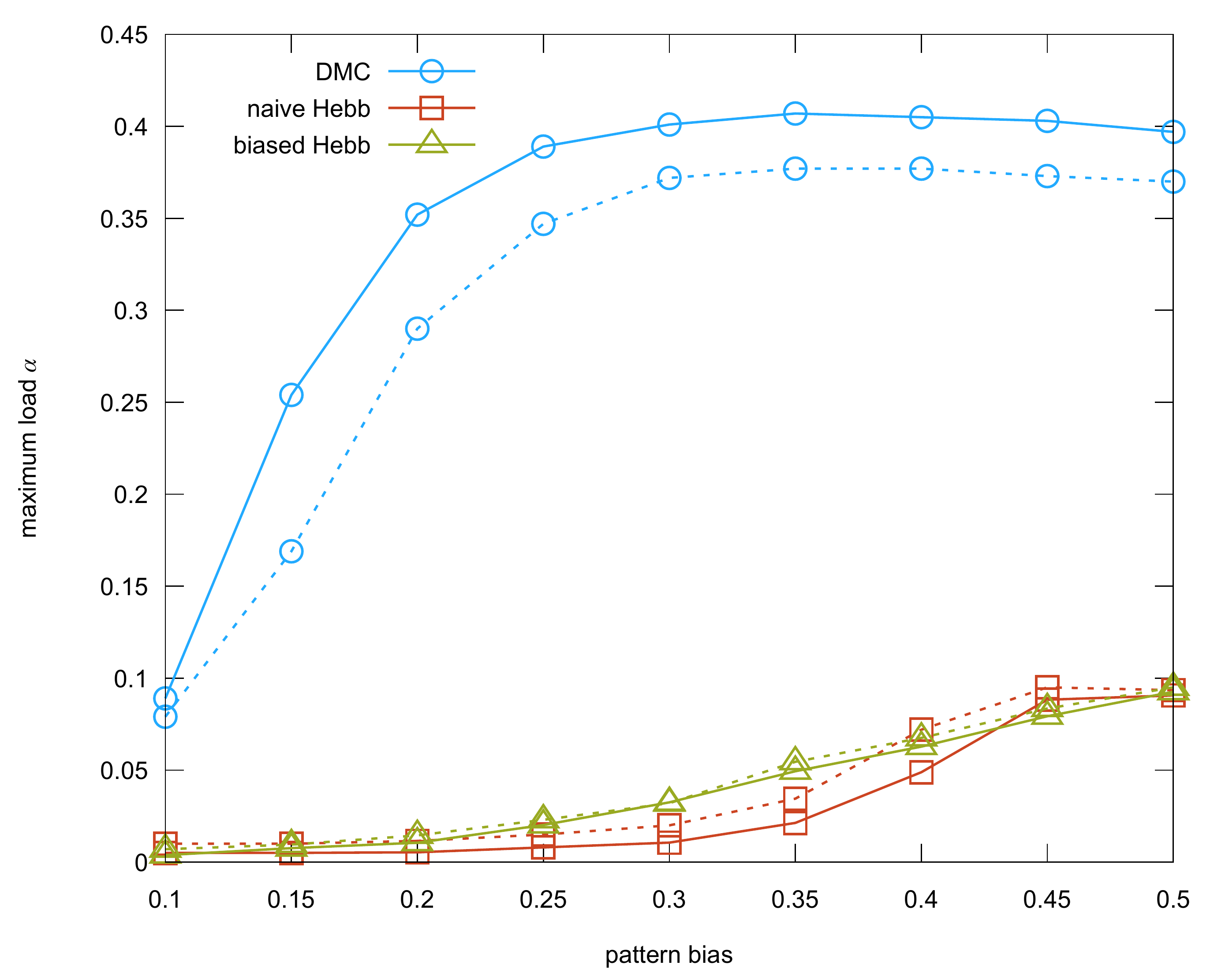}\caption{\label{fig:correlated_patts}\emph{Maximum storage load as a function
of the bias in the distribution of the patterns }for networks of size
$N=200$ (dashed curves) and $N=400$ (full curves). The correlation
is introduced trivially: each pattern is built by extracting spins
from a biased distribution $P\left(\xi_{i}\right)=b\,\delta\left(\xi_{i}-1\right)+\left(1-b\right)\delta\left(\xi_{i}+1\right)$.
The blue curves show the scaling properties of the capacity of the
DCM rule as a function of the bias. The drop in the performance for
small biases is due to finite size effects, and the performance improves
with $N$. The red and green curves shows the results for the naive
Hebb rule and the generalized Hebb rule adapted to the biased case,
respectively (see \textbf{\textbf{Appendix~\ref{subsec:correlated-patterns}}}).
For larger $N$ the capacity for all unbalanced cases is expected
to drop to $0$. All the curves were obtained by averaging over $10$
samples (error bars are small than the point size).}
\end{figure}

Most real-world data is inherently sparse and redundant, so that it
is crucial for a plasticity rule to be able to deal with a pattern
set exhibiting internal correlations. The most trivial way of introducing
a positive correlation among the patterns is to bias the probability
distribution from which the patterns are extracted, i.e. using the
probability distribution $P\left(\xi_{i}\right)=b\,\delta\left(\xi_{i}-1\right)+\left(1-b\right)\delta\left(\xi_{i}+1\right)$
for the patterns components, with $b\in\left(0,1\right)$ ($b=\nicefrac{1}{2}$
being the unbiased case). The Hebbian learning rule needs to be adapted
for enabling learning of biased patterns \citep{amit1987information_Hebb_biased}
(see \textbf{\textbf{Appendix~\ref{subsec:correlated-patterns}}}),
and the modification requires explicit knowledge of the statistics
of the stimuli. The DCM rule is instead able to adapt to the case
of unbalanced patterns without any modification, and achieves a much
better performance, as can be seen in fig.~\ref{fig:correlated_patts}. 

\begin{figure}
\begin{centering}
\includegraphics[width=0.9\textwidth]{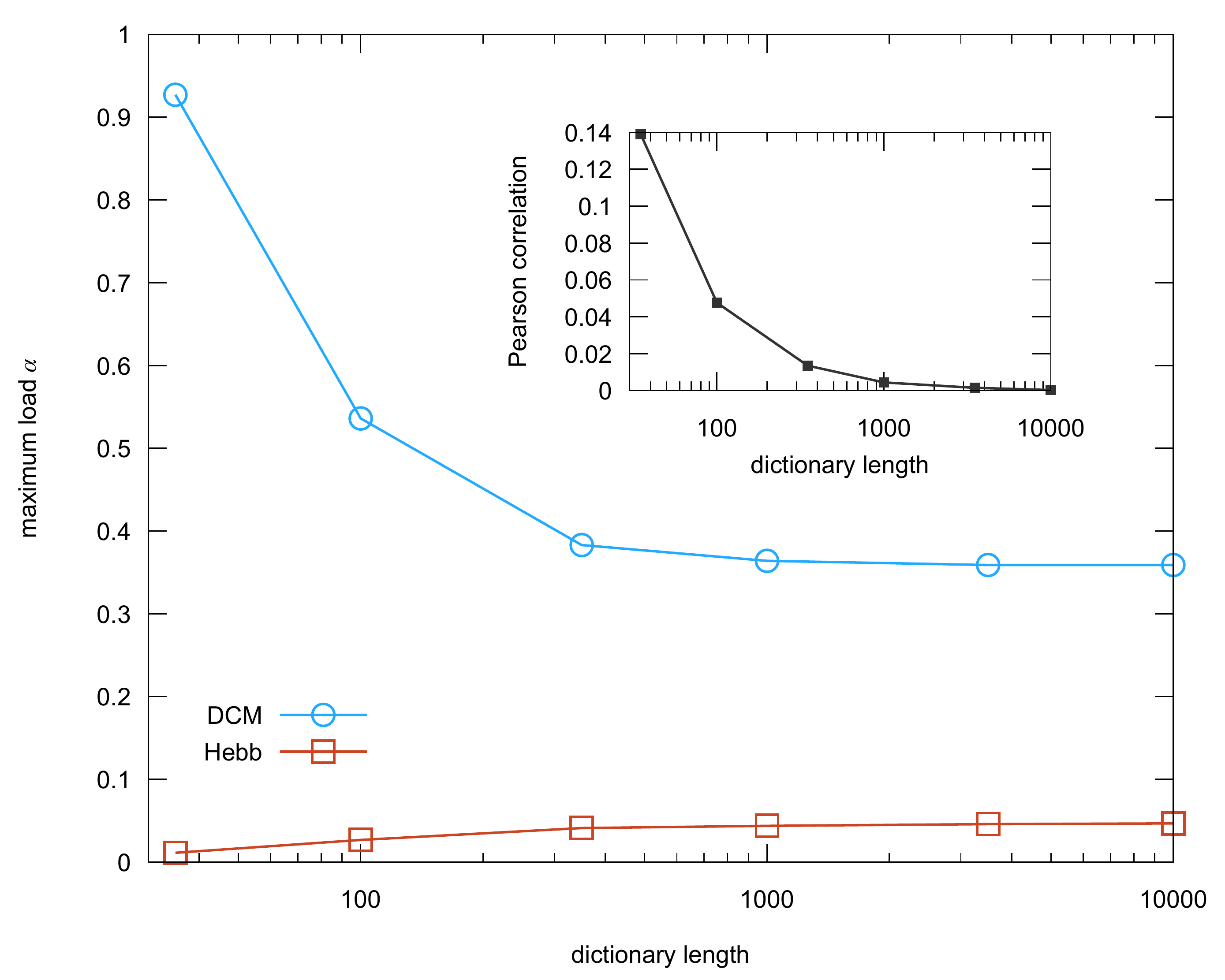}
\par\end{centering}
\caption{\emph{\label{fig:dictionary}Maximum storage load as a function of
the length of the dictionary of features }We study the critical capacity
of the generalized Hebb rule (red curve) and the DCM rule (blue curve)
when the patterns are generated as combinations of features, chosen
from a dictionary of varying length $L$. In the inset, the mean Pearson
correlation in a dataset of $200$ patterns is shown as a function
of the dictionary length. In the numerical experiments every feature
had a fixed sparsity of $f=0.1$ and each pattern was obtained as
a superposition of $F=6$ features (see \textbf{\textbf{Appendix~\ref{subsec:correlated-patterns}}}).
The curves were obtained by averaging over $10$ samples (error bars
are smaller than the point size).}
\end{figure}

A more realistic way of introducing pattern correlations can be studied
in the $s_{i}\in\left\{ 0,1\right\} $ case, where it is possible
to generate a set of patterns as combinations of sparse features drawn
from a finite length \emph{dictionary} (i.e. we pre-generate a set
of sparse patterns -- the dictionary of features $\mathcal{D}$ --
and then generate each stimulus by taking a small random subset of
$\mathcal{D}$ and superimposing the patterns within it; see \textbf{\textbf{Appendix~\ref{subsec:correlated-patterns}}}).
In the limit of an infinitely large dictionary one produces uncorrelated
patterns, but correlations set in as the length of the dictionary
is reduced. In fig.~\ref{fig:dictionary} we show how the DCM rule
is able to take advantage of the decrease in the information content
of the patterns as the total number of features is reduced.

\begin{figure}[h]
\centering{}\includegraphics[width=0.9\textwidth]{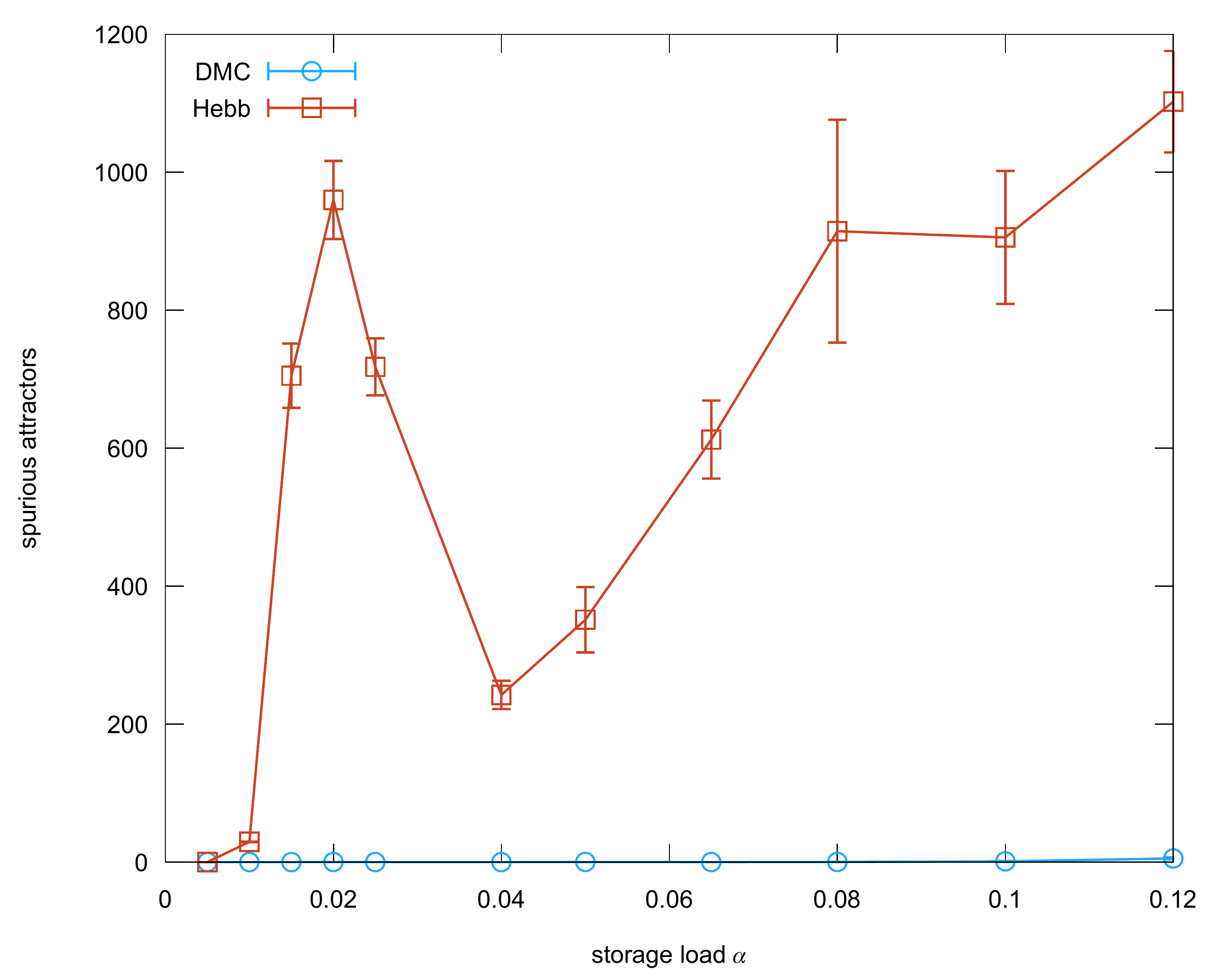}\caption{\emph{\label{fig:spurious}Number of spurious attractors} for a network
of $N=400$ neurons. This figure shows the number of distinct spurious
attractors found during $10000$ independent random walks, of $200$
time-steps, after a small number of patterns were learned by the network
(see \textbf{\textbf{Appendix~\ref{subsec:Spurious-attractors}}}).
The red curve represents the Hebb rule (the first peak is due to finite
size effects). The blue curve shows the behavior of the DCM rule.
The curves were obtained by averaging over $10$ samples (error bars
are standard errors).}
\end{figure}

Another drawback of the plain Hebb rule is the introduction of spurious
memories while the desired patterns are embedded as attractors. These
spurious states usually appear in overlapping regions of the basin
of attraction of different stored memories, and are therefore referred
as mixture states \citep{Amit:1989:MBF:77051}. As can be seen in
fig.~\ref{fig:spurious}, the problem of spurious attractors is almost
completely avoided when the DCM rule is employed, since it is able
to store the patterns more coherently and the basins of attraction
are not likely to interfere with each other. 

\subsubsection*{One-shot learning}

\begin{figure}
\begin{centering}
\includegraphics[width=0.9\textwidth]{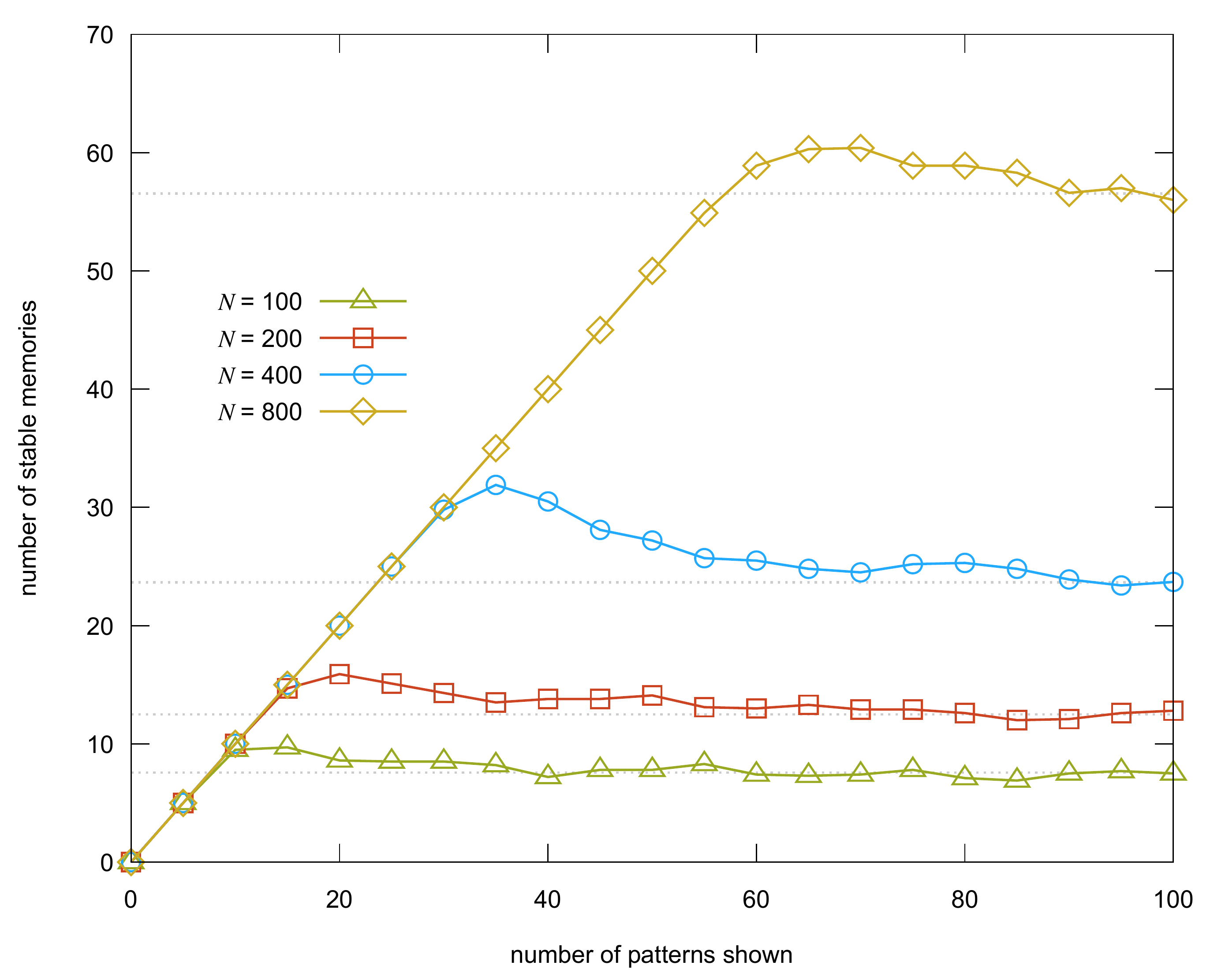}
\par\end{centering}
\caption{\emph{\label{fig:palimpsest}Scaling properties of the palimpsest
capacity }In this figure we show the results obtained when testing
the DCM learning rule in the context of one-shot learning, for the
case of $s_{i}\in\left\{ -1,+1\right\} $ neurons. The full curves
show the results for $N=100$,$N=200$,$N=400$ and $N=800$, illustrating
the scaling properties of the palimpsest capacity. The dashed gray
curves are extrapolated as the mean of the last 3 measurements. All
the points are obtained by averaging over $10$ samples.}
\end{figure}

Finally, we also tested the DCM learning rule in a one-shot on-line
setting: each pattern is presented to the network until it becomes
a stable attractor and then is never seen again. In this scenario
the relevant measure of the performance is the so called palimpsest
capacity \citep{nadal1986networks}: after an initial transient, the
network is expected to enter a steady-state regime in which an old
memory is lost every time a new one is learned. Our numerical results,
obtained in the $s_{i}\in\left\{ -1,+1\right\} $ case (fig.~\ref{fig:palimpsest}),
show that -- quite remarkably -- by simply adding a weight regularization
the DCM rule achieves an extensive palimpsest capacity, slightly above
$\sim0.05\,N$. This property was verified by a scaling analysis.
Similar results can be obtained in the $s_{i}\in\left\{ 0,1\right\} $
case only with the adaptive threshold regulatory scheme (see \textbf{\textbf{Appendix~\ref{subsec:One-shot}}}
for more details).

Another local learning rule that is known to perform well in an online
setting was proposed by Storkey \citep{storkey1998palimpsest}, and
reads:

\begin{eqnarray}
\Delta J_{ij}=\Delta J_{ji} & \propto & \left(\xi_{i}^{\mu}\xi_{j}^{\mu}-h_{i}\xi_{j}^{\mu}-h_{j}\xi_{i}^{\mu}\right)
\end{eqnarray}
where $h_{i}=\sum_{k}J_{ik}\xi_{k}$ are the local fields.\textbf{
}The last two terms can penalize the weights when the memory is already
stored ($h_{i}$ has the same sign of $\xi_{i}$) and the local field
becomes excessively large, building a regularization mechanism directly
into the learning rule. Limiting the growth of the synaptic weights
is in fact necessary in order to avoid entering a spin glass phase,
where all the memories are suddenly lost and learning can no longer
take place \citep{parisi1986memory}. However, Storkey's rule fails
when tested against our retrieval criterion in a finite temperature
setting (we are setting $\beta=2$ in the parallel Glauber dynamics).
This not only shows that the DCM is able to embed attractors arbitrarily
robustly (depending on the temperature considered during training),
but also stresses the fact that the retrieval criterion that was employed
throughout this paper is very strict compared to alternative definitions.
For example, if we consider the criterion proposed in \citep{storkey1998palimpsest}
the DCM rule palimpsest capacity is measured to be as high as $\sim0.3N$.


\subsection*{Adding hidden neuronal states}

When hidden neurons are introduced, the stochastic neural network
turns into a rather general computational device, which can be framed
as a parametric probabilistic model able to develop an internal representation
of the statistics of external stimuli. This kind of neural network
could recover a partially corrupted memory, as in an attractor neural
network, but it could also be exploited as a generative model, able
to produce new samples in accordance with the statistics inferred
from the training data.

Even in the case with undirected symmetric synaptic couplings --
the Boltzmann Machine -- the inference and learning problems become
NP-hard, since the time required for the dynamics to reach thermal
equilibrium is bound to grow exponentially with the network size \citep{salakhutdinov2008learning_Hardness_BM}.
A well studied solution to these problems is to consider a simplified
synaptic structure, in which the connections of the network are restricted
to the ones between visible and hidden neurons, the so-called Restricted
Boltzmann machine (RBM) \citep{hinton2006_RBM}. We will focus on
the same rigid architecture.

The DCM learning rule can still be understood in a KL minimization
framework. As before, in the infinite signal limit we obtain a log-pseudo-likelihood
optimization procedure, except that now the inference is from incomplete
observations and an average over all the possible hidden neuronal
states is required (see \textbf{\textbf{Appendix~\ref{sec:Visible-to-hidden}}}).
In this limit the synaptic couplings are updated as:
\begin{flushleft}
\begin{eqnarray}
\Delta J_{ij} & \propto & P\left(s_{j}|s_{\mathcal{V}}=\xi\right)\xi_{i}s_{j}-\sum_{s\in\mathcal{H}}\prod_{k\in\mathcal{H}}P\left(s_{k}|s_{\mathcal{V}}=\xi\right)P\left(s_{i}^{\prime}|s_{\mathcal{H}}\right)s_{i}^{\prime}s_{j}.\label{eq:contr_div}
\end{eqnarray}
\par\end{flushleft}

This equation is closely linked to the contrastive divergence method,
CD-$k$, a heuristic algorithm for approximating the maximum likelihood
method for RBMs \citep{carreira2005contrastive_RBM}. The first term
in eq.~(\ref{eq:contr_div}) requires sampling from the probability
distribution of the hidden neuronal state induced by a clamping stimulus
on the visible neurons, as in the positive phase of CD-$k$, while
the second term can be estimated by implementing a Gibbs sampling
chain starting from a visible state prepared in correspondence of
the stimulus but subject to no external field, as in CD-$k$'s negative
phase. This relationship could shed some light on the apparently surprising
performance that can be obtained with CD, even when a very small number
of Gibbs sampling steps $k$ is chosen: this means that the partition
function of the model is estimated very crudely, restricting sampling
only to the mode induced by the seed of the Gibbs chain. This is in
fact what the pseudo-likelihood method would require \citep{yasuda2012learning_PSL_Completed}.
CD-$k$, however, is defined in the context of models with symmetric
interactions and therefore does not apply to asymmetric kinetic models
of the type considered throughout this work.

In the presence of hidden neurons we can still apply the heuristic
prescription described above (eq.~(\ref{eq:bio_formulation})), yielding
a plasticity rule that matches time-delayed correlations, recorded
during the network dynamics. In order to test numerically how a biologically
plausible system could perform against a state-of-the-art learning
method, we also derived the Thouless-Anderson-Palmer (TAP) mean-field
equations\textbf{ }\citep{thouless1977solution_TAP} for approximating
the steady-state distribution of the neural states and the time-delayed
correlations (see \textbf{\textbf{Appendix~\ref{sec:TAP-approximation}}}
for their analytical derivation) in a sparse asymmetric network. 

In \textbf{\textbf{Appendix~\ref{sec:Visible-to-hidden}} }we consider
the problem of learning the statistics of a dataset of real-world
images \citep{lecun1998mnist}. The performance of the DCM rule is
assessed in the customary feature extraction, generative, and classification
tasks and compared with that of the TAP approach, on same neural network
architectures. While there is an obvious degradation in the learning
performance, we also observe that the robustness of our learning model
is still allowing the network to learn despite the presence of noise
and strict detrimental biological constraints. 

\section*{Discussion}

In this work we studied the problem of learning in general stochastic
neural network models. Starting from a KL divergence minimization
condition, we derived analytically a differential update rule closely
related to the Maximum Pseudo-likelihood method, able to store an
an ensemble of patterns of neuronal activity\textbf{,} conveyed to
the network in the form of external fields. With some slight modifications,
we obtained a version of the rule that allowed us to introduce a number
of important requirements for biologically plausibility, concerning
not only the network structure but the learning process as well. We
further showed that all the needed information could be collected
during the dynamics of the network by some kind of short term memory
mechanism, locally keeping track of correlations, and that the updates
could be implemented by a comparator simply trying to maintain externally
induced correlations by incrementing the synaptic weights.

Our DCM learning rule bears great resemblance with classical Hebb
plasticity, in that synaptic modifications are driven only by the
information about activity correlations locally available at the synapse.
However, the DCM rule can be applied in a general framework where
asymmetric synapses are allowed, at odds with the previous learning
paradigms. Moreover, the rule relies on finite external signals, that
are not able to quench the network dynamics completely. Apart from
retaining a higher biological plausibility, this is one of the reasons
why this rule can embed an extensive number of patterns while minimizing
the patterns cross-talk, avoiding the creation of spurious memories.
The stochastic network becomes capable of learning in a purely on-line
context, including in the extreme limit of one-shot learning.

The differential form of the plasticity rule also allows for a good
retrieval performance when the memories are correlated, both in the
case of simply biased memories and in the case of patterns obtained
as combinations of features. In the sparse case, we showed the robustness
of the DCM rule to the introduction of the excitatory-inhibitory differentiation
constraint (Dale's principle), and proposed various inhibitory mechanism
which proved to be able to control the activity level of the network
and to prevent the dynamics from reaching epileptic states.

Finally we showed how the very same learning rule allows a more general
network, in which hidden neurons are added, to perform well in feature
extraction, generation and classification tasks, when dealing with
real-world data. By means of comparison with a state-of-the art method,
we argue that, by implementing the proposed learning rule, a stochastic
neural network obeying strong biological requirements could preserve
great modeling potential. In particular, the similarities with Boltzmann
Machine learning \citep{hinton1985_BMlearning,hinton2006_RBM} (see
also below) suggest that the DCM rule may be a viable candidate for
feature extraction and inference: for example, in experiments with
patterns formed from combining features from a dictionary (as for
those of fig.~\ref{fig:dictionary}), we may hope to recover the
individual features as internal representations in the hidden part
of the network. We performed preliminary experiments in this direction
and the results are indeed promising. In this paper, however, our
numerical analysis was limited to the well-studied case of directed
visible-to-hidden synapses and digit recognition, and the exploration
of hybrid and more general architectures and tasks is left for future
work.

Future possible research directions include the generalization of
this learning framework to continuous time dynamics and more realistic
spiking network models, and the problem of learning dynamical activation
patterns instead of static ones. It must be noted that the idea of
learning recurrent weight matrix in a network model by matching some
measure of a driven system to that of an autonomous one is not new.
The general strategy for stabilizing dynamical patterns has been rediscovered
under several denominations in the broad context of reservoir computing
and generally involves the matching of local \emph{currents }\citep{Reinhart2011,Mayer2004,sussilloTransferring},
with notable examples both in the discrete time step deterministic
setting \citep{jaeger2014controlling} and in spiking network models
\citep{abbott2016building,depasquale2016using}. These models have
the advantage of capturing the dynamical complexity of neural systems.
We note that, on the other hand, they rely on some non-local learning
strategies.

Our model also shares some similarities with the Equilibrium Propagation
algorithm (EP) for energy-based models of ref.~\citep{scellier2017equilibrium},
but with some crucial differences. The main similarity relies in the
fact that the resulting update rule for the synaptic weights uses
the difference between the correlations measured with the network
in a weakly-clamped state (using the EP terminology) and a free state.
This is also reminiscent of the original algorithm for training Boltzmann
Machines proposed in ref.~\citep{hinton1985_BMlearning}. The major
difference in our model is the use of time-delayed correlations, which
stems from the different approach used in our derivation and allows
us to work in the general setting of asymmetric synaptic connections
-- indeed, the synaptic symmetry in the EP approach was regarded
by the authors as its most unsatisfactory requirement from a biological
perspective. Additional important differences arise from the overall
setting and derivation: in the EP case, the context is supervised
learning, the inputs are fully clamped and drive the network towards
an equilibrium (in the free phase), after which the outputs are weakly
clamped (the limit of vanishing clamping is considered) and the weights
updated accordingly. In our case, the context is unsupervised learning,
there is no preliminary equilibration step (the network is not reset
between pattern presentations), and the external driving force is
relatively weak but non-vanishing (it decreases to zero gradually
as training progresses).

In ref.~\citep{braunstein2011inference}, in the context of diluted
neural networks, the authors used as a learning criterion the matching
of \emph{equal-time} correlations, still comparing a system driven
by a finite field with a freely-evolving one. In that case, however,
the connections were assumed to be symmetric, and the correlations
were estimated with the Belief Propagation algorithm. At odds with
these approaches, we presented a formulation in terms of \emph{delayed
activity correlations} that, while requiring a time integration mechanism,
is completely local, and is used to construct general excitatory-inhibitory
asymmetric networks. Another attempt at devising a learning protocol
with good performances and subject to basic biological constraints
was presented in ref.~\citep{alemithreshold}, exploiting the statistics
of the inputs rather then the dynamical properties of the network.
The resulting ``Three thresholds'' learning rule (3TLR) shares with
the DCM rule most desirable features for a biological system, e.g.
it can achieve near-optimal capacity even with correlated patterns.
A detailed comparison of the performance of the two rules is technically
and computationally demanding and unfortunately out of the scope of
this work, but the 3TLR seems to require stronger driving external
fields; furthermore, lowering the field results in an abrupt performance
drop, while the DCM rule degrades gracefully, cf. fig.~\ref{fig:capacity_fields}.
\begin{acknowledgments}
C.B. and R.Z. acknowledge the Office of Naval Research Grant N00014-17-1-2569.
\end{acknowledgments}

\bibliographystyle{unsrturl}

%% file: SM.tex
\title{From statistical inference to a differential learning rule for stochastic
neural networks\\
Supplementary Material }
\author{Luca Saglietti}
\affiliation{Microsoft Research New England, Cambridge (MA), USA}
\affiliation{Italian Institute for Genomic Medicine, Torino, Italy}
\author{Federica Gerace}
\affiliation{Politecnico di Torino, DISAT, Torino, Italy}
\affiliation{Italian Institute for Genomic Medicine, Torino, Italy}
\author{Alessandro Ingrosso}
\affiliation{Center for Theoretical Neuroscience, Columbia University, New York,
USA}
\author{Carlo Baldassi}
\affiliation{Bocconi Institute for Data Science and Analytics, Bocconi University,
Milano, Italy}
\affiliation{Italian Institute for Genomic Medicine, Torino, Italy}
\affiliation{Istituto Nazionale di Fisica Nucleare, Torino, Italy}
\author{Riccardo Zecchina}
\affiliation{Bocconi Institute for Data Science and Analytics, Bocconi University,
Milano, Italy}
\affiliation{Italian Institute for Genomic Medicine, Torino, Italy}
\affiliation{International Centre for Theoretical Physics, Trieste, Italy}
\maketitle

\section{Types of neurons\label{sec:Types-of-neurons}}

In this work we considered two kinds of binary neurons, $s_{i}\in\left\{ -1,+1\right\} $
and $s_{i}\in\left\{ 0,1\right\} $. The sigmoid-shaped function $\sigma\left(\cdot\right)$
which appears in eq.~(\ref{eq:Glauber_dynamics}) takes two slightly
different forms depending on the model, as a consequence of the different
normalization term appearing in the two cases:
\begin{eqnarray}
\sigma_{\pm1}\left(s|h;\beta\right) & = & \frac{e^{\beta sh}}{e^{\beta h}+e^{-\beta h}}\label{eq:M=000026M_sigma_pm1}\\
\sigma_{01}\left(s|h;\beta\right) & = & \frac{e^{\beta sh}}{1+e^{\beta h}}.\label{eq:M=000026M_sigma_01}
\end{eqnarray}

In the case of $s_{i}\in\left\{ -1,+1\right\} $ neurons, we sampled
each component of the patterns independently from a potentially biased
probability distribution $P\left(\xi_{i}\right)=b\,\delta\left(\xi_{i}-1\right)+\left(1-b\right)\delta\left(\xi_{i}+1\right)$,
with a bias parameter $0<b<1$. In most of our tests, however, we
considered the unbiased case $b=\nicefrac{1}{2}$, except for those
presented in fig.~\ref{fig:correlated_patts}. In this case, the
local fields are naturally balanced around $0$ and the thresholds
$\theta_{i}$ can be eliminated.

In the case of $s_{i}\in\left\{ 0,1\right\} $ neurons, we sampled
each component of the memories from the prior $P\left(\xi_{i}\right)=\left(1-f_{v}\right)\,\delta\left(\xi_{i}\right)+f_{v}\delta\left(\xi_{i}-1\right)$.
Here $f_{v}$ should also correspond to the network sparsity level,
i.e. the average fraction of active neurons at a given time-step of
the network dynamics, $f_{v}=\frac{1}{N}\sum_{i=1}^{N}s_{i}$. In
this case, the thresholds $\theta_{i}$ are necessary to shift the
distribution of the local fields around zero, and we used an inhibitory
network to stabilize the overall activity (see the `Inhibitory Network
models' section below).

\section{Analytic derivation of the DCM learning rule\label{sec:Analytic-derivation}}

In this section we derive the equations for the DCM rule. For simplicity
we will consider the case of $\beta=1$. From a mathematical perspective,
we ask our learning rule to reduce the Kullback-Leibler (KL) divergence
between two different conditional probability distributions, $P\left(s^{\prime}|s;\lambda_{1}\right)$
and $P\left(s^{\prime}|s;\lambda_{2}\right)$, with $\lambda_{2}<\lambda_{1}$,
averaged over an initial state probability distribution $P\left(s\right)$.
This quantity is given by:

\begin{eqnarray}
\left\langle \mbox{KL}\left(P\left(\cdot|s;\lambda_{1}\right)||P\left(\cdot|s;\lambda_{2}\right)\right)\right\rangle _{P} & = & \sum_{s}P\left(s\right)\sum_{s^{\prime}}P\left(s^{\prime}|s;\lambda_{1}\right)\log\frac{P\left(s^{\prime}|s;\lambda_{1}\right)}{P\left(s^{\prime}|s;\lambda_{2}\right)}.\label{eq:M=000026M_KL-divergence}
\end{eqnarray}

The conditional probability is defined as a sigmoid-shaped neural
activation function (cf. eq.~(\ref{eq:Glauber_dynamics}))

\begin{equation}
P\left(s^{\prime}|s;\lambda\right)=\prod_{i=1}^{N}\sigma\left(s_{i}^{\prime}|h_{i}^{\lambda}\right),\label{eq:M=000026M_conditional_probability}
\end{equation}
with local fields $h_{i}^{\lambda}$ given by: $h_{i}^{\lambda}=h_{i}^{\mathrm{ext},\lambda}+\sum_{j\neq i}J_{ij}^{\lambda}s_{j}-\theta_{i}^{\lambda}$.
Here we adopt the superscript $\lambda$ to distinguish between the
two networks, subject to different external field intensities $\lambda$. 

Plugging the expression~(\ref{eq:M=000026M_conditional_probability})
for the conditional probability into the definition~(\ref{eq:M=000026M_KL-divergence})
of the KL divergence, we exploit the factorization property of the
single neuron conditional probabilities, in order to isolate the $i$-th
contribution and trace out all the others. Therefore, we get the final
expression for the averaged KL divergence: 

\begin{eqnarray}
\left\langle \mbox{KL}\left(P\left(\cdot|s;\lambda_{1}\right)||P\left(\cdot|s;\lambda_{2}\right)\right)\right\rangle _{P} & = & \sum_{s}P\left(s\right)\sum_{i}\sum_{s_{i}^{\prime}}\sigma\left(s_{i}^{\prime}|h_{i}^{\lambda_{1}}\right)\log\frac{\sigma\left(s_{i}^{\prime}|h_{i}^{\lambda_{1}}\right)}{\sigma\left(s_{i}^{\prime}|h_{i}^{\lambda_{2}}\right)}
\end{eqnarray}

The next step is to minimize this quantity, by differentiating with
respect to $J_{ik}^{\lambda_{2}}$ and $\theta_{i}^{\lambda_{2}}$,
asking the second network to compensate for the decrease in the external
field through an adaptation of its parameters. For both expressions
of $\sigma$ of eqs.~(\ref{eq:M=000026M_sigma_pm1}) and~(\ref{eq:M=000026M_sigma_01}),
the following property holds:
\begin{equation}
\frac{1}{\beta}\frac{\partial}{\partial h}\log\sigma\left(s|h;\beta\right)=s-\left\langle s\right\rangle _{h}
\end{equation}
where here $\left\langle s\right\rangle _{h}=\sum_{s}s\,\sigma\left(s|h;\beta\right)$.
This allows us to derive the following simple formulas for the derivatives
with respect to the parameters, for both neuronal models:

\begin{eqnarray}
-\frac{1}{\beta}\frac{\partial}{\partial J_{ik}^{\lambda_{2}}}\left\langle \mbox{KL}\left(P\left(\cdot|s;\lambda_{1}\right)||P\left(\cdot|s;\lambda_{2}\right)\right)\right\rangle _{P} & = & \sum_{s}P\left(s\right)\sum_{s_{i}^{\prime}}\sigma\left(s_{i}^{\prime}|h_{i}^{\lambda_{1}}\right)\left(s_{i}^{\prime}-\left\langle s_{i}^{\prime}\right\rangle \right)s_{k}\nonumber \\
 & = & \left\langle s_{i}^{\prime}s_{k}\right\rangle _{P,\lambda_{1}}-\left\langle s_{i}^{\prime}s_{k}\right\rangle _{P,\lambda_{2}}\label{eq:M=000026M_grad_Jik}\\
-\frac{1}{\beta}\frac{\partial}{\partial\theta_{i}^{\lambda_{2}}}\left\langle \mbox{KL}\left(P\left(\cdot|s;\lambda_{1}\right)||P\left(\cdot|s;\lambda_{2}\right)\right)\right\rangle _{P} & = & \sum_{s}P\left(s\right)\sum_{s_{i}^{\prime}}\sigma\left(s_{i}^{\prime}|h_{i}^{\lambda_{1}}\right)\left(s_{i}^{\prime}-\left\langle s_{i}^{\prime}\right\rangle \right)\nonumber \\
 & = & -\left(\left\langle s_{i}^{\prime}\right\rangle _{P,\lambda_{1}}-\left\langle s_{i}^{\prime}\right\rangle _{P,\lambda_{2}}\right)
\end{eqnarray}
As mentioned above though, the second one is not actually used in
the $\pm1$ model since we did not use the thresholds $\theta_{i}$
in that case.

\subsection{Connection with maximum pseudo-likelihood method\label{subsec:pseudo-likelihood}}

In the fully visible case, the clamped probability distribution eq.~(\ref{eq:clamped})
simply becomes $P_{\mathrm{clamp}}\left(s;\xi\right)=\prod_{i=1}^{N}\delta_{s_{i},\xi_{i}}$,
and the average KL divergence defined in eq.~(\ref{eq:M=000026M_KL-divergence})
can be written explicitly as:

\begin{eqnarray}
 &  & \left\langle KL\left[P\left(\cdot|s;\lambda^{\mathrm{ext}}=\infty\right)||P\left(\cdot|s;\lambda^{\mathrm{ext}}=0\right)\right]\right\rangle _{P_{\mathrm{clamp}}\left(\xi\right)}=\label{eq:M=000026M_psl-attractors}\\
 &  & =-\sum_{i=1}^{N}\log P\left(s_{i}=\xi_{i}|\left\{ s_{j}=\xi_{j}\right\} _{j\neq i};\lambda^{\mathrm{ext}}=0\right).\nonumber 
\end{eqnarray}
This expression can be recognized as one of the terms appearing in
the so called log-pseudo-likelihood $\mathcal{L}\left(\left\{ \xi^{\mu}\right\} |J_{ij},\theta;\beta\right)=\frac{1}{M}\sum_{\mu=1}^{M}\sum_{i=1}^{N}\log P\left(s_{i}=\xi_{i}^{\mu}|\left\{ s_{j}=\xi_{j}^{\mu}\right\} _{j\neq i};\lambda^{ext}=0\right)$. 

The pseudo-likelihood method provides a computationally inexpensive
yet statistically consistent estimator \citep{gidas1988consistency_ML_PML}
when the functional form of the joint probability distribution over
the configurations is unknown, and is thus approximated in the factorized
form $P\left(s=\xi^{\mu}\right)=\prod_{i}P\left(s_{i}=\xi_{i}^{\mu}|\left\{ s_{j}=\xi_{j}^{\mu}\right\} _{j\neq i}\right)$.
In the framework of learning, the minimization of eq.~(\ref{eq:M=000026M_psl-attractors})
can be seen instead as a stability requirement for the memory $\xi$,
as it progressively increases the probability that the stochastic
dynamics remains fixed in the attractor state. 

\subsection{Connection with the perceptron rule\label{subsec:perceptron}}

In the noise-free limit $\beta\to\infty$, where the state of the
neuron $s_{i}^{t+1}$ is deterministically obtained by taking the
sign of the local incoming current, the pseudo-likelihood synaptic
weight update would read:

\begin{eqnarray}
\Delta J_{ij} & = & \begin{cases}
0 & \xi_{i}h_{i}\ge0\\
2\eta\xi_{i}^{\mu}\xi_{j}^{\mu} & \xi_{i}h_{i}<0
\end{cases},
\end{eqnarray}
which is the well-known perceptron rule. Indeed, since the next state
of a neuron is conditionally dependent on the previous state of the
other $N-1$ neurons, one can reinterpret the problem of learning
a certain number of attractors as $N$ independent perceptron learning
problems. In a zero temperature setting, the incoming weights of a
neuron $i$ can be simply updated whenever its predicted state is
misaligned with respect to the $i$-th component of the memory to
be learned, $s_{i}^{t+1}\neq\xi_{i}$, by shifting its weights in
the direction of the desired state and in parallel to the pattern
itself. It is known that the perceptron rule saturates the theoretical
Gardner bound $\alpha_{c}=2$ for the critical memory capacity of
a fully-visible neural network at zero noise \citep{gardnerSpace}. 

Moreover, if we follow \citep{baldassi2018inverse} and consider negative
field intensities $\lambda_{\text{min}}<0$ (instead of $\lambda_{\text{min}}=0$
as in the pseudo-likelihood method), we obtain:

\begin{eqnarray}
\Delta J_{ij} & = & \begin{cases}
0 & \xi_{i}h_{i}\ge\left|\lambda_{\text{min}}\right|\\
2\eta\xi_{i}^{\mu}\xi_{j}^{\mu} & \xi_{i}h_{i}<\left|\lambda_{\text{min}}\right|
\end{cases}.
\end{eqnarray}
This is nothing but the perceptron rule with robustness parameter
$\left|\lambda_{\text{min}}\right|$, that forces the network to learn
the memories so that they are attractive in a full sphere of such
radius. However, any $\lambda_{\text{min}}<0$ will also cause the
maximum capacity of the network to decrease \citep{engel2001statistical}.

\section{Inhibitory Network models\label{sec:Inhibitory-Network-models}}

We considered three different schemes that can reproduce the effect
of an inhibitory network. In the first one, the inhibitory network
is replaced by a global inhibitory unit connected to all the $N$
excitatory neurons \citep{alemithreshold}, which elastically drives
the system towards the desired activity level through a feed-back
signal. An alternative scheme can be obtained by introducing a soft
``winner-takes-all'' mechanism, effectively playing the role of
a global inhibitory unit \citep{binas2014learning,douglas1989canonical,mountcastle1997columnar,binzegger2004quantitative,douglas2007recurrent,carandini2012normalization,handrich2009biologically,mao2007dynamics,lynch2016computational,oster2006spiking,fang1996dynamics}.
This mechanism is meant to model a continuous time scale phenomenon:
the neurons with higher local activities could become active before
the others and start to excite the inhibitory component of the network,
whose feed-back signal is triggered when the correct fraction $f_{v}$
of neurons is already active; this signal thus depresses all the local
activities of the network, preventing the remaining neurons from activating.
The last inhibitory scheme is based on the introduction of locally
adaptive thresholds (from a biological point of view, this mechanism
can be justified with the widely observed phenomenon of thresholds
variability in the central nervous system \citep{fontaine2014spike}). 

The aim of the inhibitory feedback is to maintain the excitatory network
activity around a desired level, preventing epileptic (all-on) or
completely switched off states in the $\left\{ 0,1\right\} $ model.
In the following, we provide more detailed explanations and some implementation
details for each scheme. 

\subsection{The global inhibitory unit scheme}

We consider a generalization of the global inhibitory unit scheme
proposed in~\citep{alemithreshold}, for a purely excitatory stochastic
neural network constituted by an ensemble of $N$ neurons. Suppose
that, within the entire neuronal population, we can distinguish $G$
different groups of neurons, such that $N=\sum_{\alpha=1}^{G}N_{\alpha}$,
with different sparsity levels. We introduce $G$ global inhibitory
units, whose task is to maintain the activity $S^{\alpha}=\sum_{i=1}^{N}s_{i}^{\alpha}$
of each population of neurons at the desired level $f^{\alpha}N_{\alpha}$.
According to the global inhibitory unit scheme, each excitatory neuronal
ensemble $\alpha$ receives a feed-back signal $\mathcal{I}^{\alpha}\left(\left\{ f^{\beta},S_{\beta}\right\} _{\beta=1}^{G}\right)$,
which can be parametrized as:

\begin{eqnarray}
\mathcal{I}^{\alpha}\left(\left\{ f^{\beta},S_{\beta}\right\} _{\beta=1}^{G}\right) & = & H_{0}^{\alpha}+\nu^{\alpha\alpha}\left(S_{\alpha}-f^{\alpha}N_{\alpha}\right)+\sum_{\beta\neq\alpha}\nu^{\alpha\beta}\left(S_{\beta}-f^{\beta}N_{\beta}\right).
\end{eqnarray}
In this section we derive analytically an expression for both the
global inhibition constant $H_{0}^{\alpha}$ and the parameters $\nu^{\alpha\beta}$
that control the elastic reaction to possible oscillations around
the desired activities. 

Assuming that the local fields $h_{i}^{\alpha}$ in population $\alpha$
are Gaussian distributed, the inhibitory units are required to correctly
set the mean of the distribution around the mean threshold $T^{\alpha}=\left\langle \theta_{i}^{\alpha}\right\rangle $,
so that the integral of the distribution above threshold contains
exactly $f^{\alpha}N_{\alpha}$ local fields:

\begin{eqnarray}
\left\langle h_{i}^{\alpha}\right\rangle  & = & T^{\alpha}-H^{-1}\left(f^{\alpha}\right)\sigma_{\alpha}.\label{eq:M=000026M_mlocalfield_shifted}
\end{eqnarray}
Here $H^{-1}\left(x\right)=\sqrt{2}\mbox{erfc}^{-1}\left(2x\right)$
represents an inverse error function, determining the proper shift
to be applied, measured in units of the standard deviation of the
distribution $\sigma_{\alpha}$. The latter can be easily computed,
giving:
\begin{eqnarray}
\sigma_{\alpha} & = & \sqrt{\left(\sigma_{J}^{\alpha\alpha}\right)^{2}\left(S^{\alpha}-f^{\alpha}\right)+\sum_{\beta\neq\alpha}\left(\sigma_{J}^{\alpha\beta}\right)^{2}S^{\beta}},
\end{eqnarray}
where $\sigma_{J}^{\alpha\beta}$ stands for the standard deviation
of the distribution of the synaptic couplings from population $\beta$
to population $\alpha$.

By summing and subtracting $\sum_{\beta}\left(\sigma_{J}^{\alpha\beta}\right)^{2}f^{\beta}N_{\beta}$
in the square root, assuming small deviations of the activity of the
network $S^{\alpha}$ from the desired activity level $f^{\alpha}N_{\alpha}$,
we can expand $\sigma_{\alpha}$ obtaining:
\begin{eqnarray}
\sigma_{\alpha} & = & \sqrt{f^{\alpha}\left(N_{\alpha}-1\right)\left(\sigma_{J}^{\alpha\alpha}\right)^{2}+\sum_{\beta\neq\alpha}\left(\sigma_{J}^{\alpha\beta}\right)^{2}f^{\beta}N_{\beta}}\times\\
 &  & \times\left(1+\frac{\left(\sigma_{J}^{\alpha\alpha}\right)^{2}\left(S^{\alpha}-f^{\alpha}N_{\alpha}\right)+\sum_{\beta\neq\alpha}\left(\sigma_{J}^{\alpha\beta}\right)^{2}\left(S^{\beta}-f^{\beta}N_{\beta}\right)}{2\left(f^{\alpha}\left(N_{\alpha}-1\right)\left(\sigma_{J}^{\alpha\alpha}\right)^{2}+\sum_{\beta\neq\alpha}\left(\sigma_{J}^{\alpha\beta}\right)^{2}f^{\beta}N_{\beta}\right)}\right).\nonumber 
\end{eqnarray}

In the left hand side of eq.~(\ref{eq:M=000026M_mlocalfield_shifted}),
instead, each local field $h_{i}^{\alpha}$ is given by the sum of
three different contributions, namely the external field, the recurrent
input and the feed-back signal from the inhibitory unit:
\begin{eqnarray}
\left\langle h_{i}^{\alpha}\right\rangle  & = & \left<h_{i}^{\mathrm{ext},\alpha}+\sum_{j\neq i}J_{ij}^{\alpha\alpha}s_{j}^{\alpha}+\sum_{\beta\neq\alpha}\sum_{j}J_{ij}^{\alpha\beta}s_{j}^{\beta}-H_{0}^{\alpha}+\right.\\
 &  & \left.\quad-\nu^{\alpha\alpha}\left(S^{\alpha}-f^{\alpha}N^{\alpha}\right)-\sum_{\beta\neq\alpha}\nu^{\alpha\beta}\left(S^{\beta}-f^{\beta}N_{\beta}\right)\right>.\nonumber 
\end{eqnarray}
We can compute the average by summing and subtracting $\sum_{\beta}\overline{J^{\alpha\beta}}S^{\beta}$,
obtaining:
\begin{eqnarray}
\left\langle h_{i}^{\alpha}\right\rangle  & = & \overline{h^{\mathrm{ext},\alpha}}+\overline{J^{\alpha\alpha}}\left(S^{\alpha}-f^{\alpha}\right)-\sum_{\beta\neq\alpha}\overline{J^{\alpha\beta}}S^{\beta}-H_{0}^{\alpha}-\nu^{\alpha\alpha}\left(S^{\alpha}-f^{\alpha}N^{\alpha}\right)+\\
 &  & -\sum_{\beta\neq\alpha}\nu^{\alpha\beta}\left(S^{\beta}-f^{\beta}N_{\beta}\right).\nonumber 
\end{eqnarray}

We therefore get an expression for the global inhibitory constant
$H_{0}^{\alpha}$ and the parameters $\nu^{\alpha\alpha}$ and $\nu^{\alpha\beta}$
that satisfy eq.~(\ref{eq:M=000026M_mlocalfield_shifted}):

\begin{eqnarray}
H_{0}^{\alpha} & = & \overline{h^{\mathrm{ext},\alpha}}+\left(N_{\alpha}-1\right)\overline{J^{\alpha\alpha}}f^{\alpha}+\sum_{\beta\neq\alpha}N_{\beta}\overline{J^{\alpha\beta}}f^{\beta}+\\
 &  & +\,\,H^{-1}\left(f^{\alpha}\right)\sqrt{f^{\alpha}\left(N_{\alpha}-1\right)\left(\sigma_{J}^{\alpha\alpha}\right)^{2}+\sum_{\beta\neq\alpha}\left(\sigma_{J}^{\alpha\beta}\right)^{2}f^{\beta}N_{\beta}}-T^{\alpha}\nonumber \\
\nu^{\alpha\alpha} & = & \overline{J^{\alpha\alpha}}+\frac{H^{-1}\left(f^{\alpha}\right)\left(\sigma_{J}^{\alpha\alpha}\right)^{2}}{2\sqrt{f^{\alpha}\left(N_{\alpha}-1\right)\left(\sigma_{J}^{\alpha\alpha}\right)^{2}+\sum_{\beta\neq\alpha}\left(\sigma_{J}^{\alpha\beta}\right)^{2}f^{\beta}N_{\beta}}}\\
\nu^{\alpha\beta} & = & \overline{J^{\alpha\beta}}+\frac{H^{-1}\left(f^{\alpha}\right)\left(\sigma_{J}^{\alpha\beta}\right)^{2}}{2\sqrt{f^{\alpha}\left(N_{\alpha}-1\right)\left(\sigma_{J}^{\alpha\alpha}\right)^{2}+\sum_{\beta\neq\alpha}\left(\sigma_{J}^{\alpha\beta}\right)^{2}f^{\beta}N_{\beta}}}.
\end{eqnarray}

Notice that a contribution to the global inhibitory constant $H_{0}^{\alpha}$
arises from the mean external field $\overline{h^{\mathrm{ext},\alpha}}=\lambda^{\mathrm{ext}}f^{\alpha}$,
so that neurons that do not receive an excitatory external stimulus
are effectively depressed. Since the adaptation of the synaptic couplings
according to the plasticity rule is considered to be adiabatic, the
means and the standard deviations required for setting a correct inhibition
are affected only over longer time scales and need not be updated
instantaneously.

The scheme described here can be easily specialized to the simple
cases of fully visible or visible-to-hidden restricted connectivity,
which have been analyzed in detail in this work.

\subsection{Soft ``winner takes all'' mechanism}

This inhibitory scheme can be easily implemented in the synchronous
dynamics considered in this work: before the new neuronal state gets
extracted (eq.~(\ref{eq:Glauber_dynamics})), the local activities
are first sorted with respect to their magnitude, then a global inhibitory
input is added, whose value is set just below the activation of the
$\left(fN\right)$-th highest excited neuron. This procedure guarantees
a fine-tuned control on the sparsity level $f$ of the network. When
the network is composed of a number $G$ of different groups of neurons,
each with a different sparsity level, the sorting operation is done
inside each group. Some theoretical results show that neurons with
adaptive threshold perform better than those with a constant threshold
in presence of highly correlated stimuli \citep{huang2016adaptive}:
we confirm these observations, since we have seen that this scheme
is the best one in the one-shot learning task.

\subsection{The adaptive thresholds regulatory scheme}

The $s_{i}\in\left\{ 0,1\right\} $ case can be mapped exactly on
the $s_{i}\in\left\{ -1,+1\right\} $ case, but this operation requires
the thresholds to dynamically adapt to any change in the synaptic
couplings.

In order to obtain the correct mapping one can consider the conditional
probabilities of the two models, and look for a transformation of
the neural variables and of the parameters which allows to move between
the two scenarios. After inserting the simple change of variables
$s_{i}\to s_{i}^{\prime}=\frac{\left(s_{i}+1\right)}{2}$ in the expression
for the local activities in the $s_{i}\in\left\{ -1,+1\right\} $
model (note that in this section the $s^{\prime}$ notation is \emph{not}
used to denote the next step of the dynamics), we get the matching
equation: 

\begin{equation}
h_{i}^{\prime}=2\left(h_{i}^{\mathrm{ext}}+\sum_{j}J_{ij}\left(2s_{i}^{\prime}-1\right)\right)=\left.h_{i}^{\mathrm{ext}}\right.^{\prime}+\sum_{j}J_{ij}^{\prime}s_{i}^{\prime}-\theta_{i}^{\prime}.
\end{equation}
which is satisfied by posing $\left.h_{i}^{\mathrm{ext}}\right.^{\prime}=2h_{i}^{\mathrm{ext}}$,
$J_{ij}^{\prime}=4J_{ij}$ and $\theta_{i}^{\prime}=2\sum_{j}J_{ij}=\frac{1}{2}\sum_{j}J_{ij}^{\prime}$
in the case of $s_{i}\in\left\{ 0,1\right\} $ neurons.

By looking at the case with $f_{v}=0.5$ sparsity, this mapping suggests
that it is possible to set the thresholds in correspondence of the
average value of the incoming excitatory stimuli received by each
neuron:

\begin{eqnarray}
\theta_{i} & = & \left\langle \sum_{j\neq i}J_{ij}s_{j}^{t}\right\rangle _{t}=f_{v}\sum_{j\neq i}J_{ij}.
\end{eqnarray}

This definition properly matches the one obtained from the exact mapping
just in the case of $f_{v}=0.5$. However, this choice was found to
allow an extensive capacity in the on-line learning regime for the
$s_{i}\in\left\{ 0,1\right\} $ neuronal state variables even with
different sparsity levels. Having set the thresholds in such a way,
one gets a slightly different form for the local activations $h_{i}=h_{i}^{\mathrm{ext}}+\sum_{j\neq i}J_{ij}\left(s_{j}^{t}-f_{v}\right)$
and the learning rule eq.~(\ref{eq:M=000026M_grad_Jik}) changes
to:

\begin{eqnarray}
\Delta J_{ij} & \propto & \left(\left\langle s_{i}^{t+1}\left(s_{j}^{t}-f_{v}\right)\right\rangle _{t,\lambda_{1}}-\left\langle s_{i}^{t+1}\left(s_{j}^{t}-f_{v}\right)\right\rangle _{t,\lambda_{2}}\right).
\end{eqnarray}

\section{Simulation: Implementation details\label{sec:Simulation}}

We provide here detailed description of the learning algorithm in
its heuristic version, described at the end of 'The Model' section,
with the update rule eq.~(\ref{eq:bio_formulation}). %
\begin{comment}
Suppose the network size $N$ is fixed and that a set of patterns
$\left\{ \xi^{\mu}\right\} _{\mu=1}^{M}$, to be presented to the
visible neurons of the network, is given. 
\end{comment}
The learning protocol consists of an iterative optimization procedure
where the parameters $J$ and $\theta$ are incrementally updated.
Throughout this work we initialized the weights $J$ uniformly at
random; for the $\pm1$ models, they were sampled from the interval
$\left[-\frac{1}{\sqrt{N}},\frac{1}{\sqrt{N}}\right]$, while for
the $0/1$ models they were sampled from the interval $\left[0,\frac{1}{\sqrt{N}}\right]$.
The thresholds $\theta$ were set to $0$ in the $\pm1$ model, and
initialized all to the same value in the $0/1$ model (the precise
value is essentially irrelevant because of the effect of the inhibitory
network; we used $0.35$ in our simulations).

Every pattern $\xi^{\mu}$ is presented in the form of an external
field $h^{\mathrm{ext}}=\lambda^{\mathrm{ext}}\xi^{\mu}$, where the
signal intensity is initialized at a fixed value $\lambda_{\max}$
and then progressively decreased to zero in steps of $\Delta\lambda$.
Before the learning process starts, we let the network evolve towards
a state correlated with the pattern by waiting for a few iterations
$T_{\mathrm{init}}$ of the dynamics while the external field set
to its maximum. Then, the learning process starts in the ``positive''
phase, registering the correlations in a time window of $T$ steps
at an initial value of the external field $\lambda$, and subsequently
at a value lowered by $\Delta\lambda$, in the ``negative'' phase.
The parameters are updated with a fixed learning rate $\eta$, as
in eq.~(\ref{eq:bio_formulation}). 

This procedure is repeated until the external field reaches zero.
The length of the time windows $T$ has to be chosen in such a way
that the state reached at the end of each averaging procedure is still
in nearly the same region around the pattern, otherwise another initialization
phase would be needed. In our experiments, we found that a good performance
is achieved when $T$ ranges from $\sim3$ to $\sim25$, provided
the learning rate is lowered when the average is taken over very few
iterations. This shows that a network implementing the DCM plasticity
rule is able to learn even in the presence of an extremely low signal-to-noise
ratio.

The relevant computation can be parallelized, since all the quantities
involved (both in the dynamics and in the learning process) are local
with respect to the synapses and the neurons. A simple pseudo-code
implementation scheme for the learning protocol can be found in algorithm~\ref{alg:learning_algo-1}.

The learning rate is constant in time and was arbitrarily set to $\eta=0.01$
in our simulations. 

\begin{algorithm}[h]
\begin{raggedright}
\KwIn{parameters: $\eta$, $cycles$, $\lambda_{max}$, $\Delta \lambda$,  $T$, $T_\mathrm{init}$} 
Initialize $J$ randomly $\sim U\left(\frac{-1}{\sqrt{N}},\frac{1}{\sqrt{N}}\right)$\;
\For{$\mbox{cycle} = 1$ \KwTo cycles}{
\For{$\mu$ in random permutation of $[1:p]$ }{
Set the external field on the visible neurons to an intensity $\lambda_{max}$\;
Run the network for $T_\mathrm{init}$ steps\; 
\While{$\lambda > 0 $}{
Estimate $\left\langle s_{i}^{t+1} s_{j}^{t} \right\rangle_{\lambda}$ for $T$ steps\;
Estimate $\left\langle s_{i}^{t+1} s_{j}^{t} \right\rangle_{\lambda- \Delta \lambda}$ for $T$ steps\;
$J_{ij} \leftarrow J_{ij} + \eta \left[ \left\langle s_{i}^{t+1} s_{j}^{t} \right\rangle_{\lambda} - \left\langle s_{i}^{t+1} s_{j}^{t} \right\rangle_{\lambda - \Delta \lambda}\right]$\;
$\lambda \leftarrow \lambda - \Delta \lambda$\;
}}
\If{all patterns are learned}{BREAK\;}
}
\ 
\par\end{raggedright}
\caption{\label{alg:learning_algo-1}Pseudo-code implementation scheme for
the DCM learning protocol (fig.~\ref{fig:protocol}). For simplicity,
we report the scheme used for $\pm1$ network models.}
\end{algorithm}

\subsection{Measuring the width of the basins of attraction\label{subsec:Basins-width}}

We introduced an operative measure of the basin size, relating it
to the level of corruption of the memories before the retrieval: a
set of $M=\alpha N$ patterns is considered to be successfully stored
at a noise level $\chi$ if, initializing the dynamics in a state
where a fraction $\chi$ of the pattern is randomly corrupted, the
retrieval rate for each pattern is at least $90\%$ (as estimated
from $100$ separate trials per pattern) after at most $1000$ learning
cycles ($250$ in the simulations with finite fields). A successful
retrieval is measured when, in absence of external input, the network
evolves towards a neuronal state with an overlap $\ge0.99$ with the
learned pattern in at most $50$ steps of the dynamics.

\subsection{Spurious attractors\label{subsec:Spurious-attractors}}

In the numerical experiments for fig.~\ref{fig:spurious}, the storage
load $\alpha$ was chosen to be sufficiently small, such that both
the DCM rule and the Hebb rule are able to learn stable attractors.
The presence of spurious attractors was detected as follows: the network
state was initialized at random, and was then allowed to evolve freely
for $200$ time steps. After this initialization period, the magnetization
was recorded for a few iterations and compared with the stored attractors.
If the modulus of the overlap with any one of them was $>0.95$, the
state was considered to have reached a known attractor. Otherwise,
the magnetization was recorded for some more iterations, in order
to check if a stable state was reached, and if this condition occurred
the magnetization was clipped to $\pm1$ and a new spurious attractor
was counted. In the following random restart, this attractor was inserted
in the list of known attractors. Of course, this procedure only provides
an estimate of the number of distinct spurious attractors introduced
by the learning rule, but sufficient to highlight a large, qualitative
difference in the behavior of DCM compared to the Hebb rule.

The first peak in fig.~\ref{fig:spurious}, in the Hebb's curve plot,
is due to finite size effects: the number of spurious states is expected
to grow at least exponentially in a sub-extensive regime $M\ll N$.
In the extensive regime, mixtures of odd number of memories can still
be observed, but as the storage load $\alpha$ is increased the mixture
states composed of larger number of patterns are expected to disappear,
and the growth in the number of spurious attractors is no longer exponential
\citep{amit1987statistical}.

\subsection{One-shot tests and palimpsest regime\label{subsec:One-shot}}

In the one-shot simulations every pattern is seen by the network only
once, and its memory will eventually be overwritten by the new ones.
The goal is to reach a steady state regime in which, at each new presentation,
the last $M$ learned memories maintain the required stability. This
storage load is called the palimpsest capacity.

In order to reach the maximal capacity, the parameters have to be
fine tuned so that the learning process for each memory is slow and
the most recently learned ones are minimally perturbed: one has to
ensure that in the freely-evolving dynamics, i.e. the last time window
during the external field pulse, the neuronal state does not escape
the basin of attraction of the new memory and enters a previously
learned one, causing the loss of existing memories. In the simulations
presented in fig.~\ref{fig:palimpsest} we addressed this problem
rather drastically by simply removing this last window, which only
resulted in a slight improvement in the palimpsest capacity.

We also set $\eta=0.01$, $\lambda_{max}=4$, $\Delta\lambda=1$ and
the length of the time windows was chosen to be slightly faster then
in the other tests, $T=10$. In this setting the number of presentations
of the same pattern, i.e. the number of external field pulses, required
for reaching its desired stability is around $\sim1000$. This number
would grow in time, because of the increase in the average connectivity
of the network as new memories are added, a problem that can be overcome
with the introduction of a synaptic weight $L2$-regularization.

In the case of $s_{i}\in\left\{ 0,1\right\} $ neurons, we surprisingly
lose the property of an extensive palimpsest capacity. The problem
seems to be related to the need for a substantial shift in the threshold,
that would allow a wide basin of attraction for a new pattern, as
expressed by eq.~(\ref{eq:bio_formulation}). This modification seems
to strongly affect the network dynamics also when it hovers around
a different, previously learned memory, introducing a disruptive effect
in the palimpsest regime. In the normal learning task, instead, the
thresholds are eventually set to a level which is compatible with
all the patterns, since the learning protocol can cycle through the
pattern set many times. The only way we found to obtain a good performance
in the one-shot learning task for $s_{i}\in\left\{ 0,1\right\} $
neurons with our model is to introduce an adaptive threshold regulatory
scheme, stemming from a direct mapping to the $s_{i}\in\left\{ -1,+1\right\} $
case.

\subsection{Generation of correlated patterns\label{subsec:correlated-patterns}}

In the case of $s_{i}\in\left\{ -1,+1\right\} $ neurons, we only
introduced correlations in the form of a bias in the generation of
the patterns, see section `Types of neurons' above. Note that, in
the biased case $b\ne\nicefrac{1}{2}$, it is known that the naive
Hebb rule $J_{ij}=\frac{1}{M}\sum_{\mu}\xi_{i}^{\mu}\xi_{j}^{\mu}$
has to be generalized to $J_{ij}=\frac{1}{M}\sum_{\mu}\left(\xi_{i}^{\mu}-2b+1\right)\left(\xi_{j}^{\mu}-2b+1\right)$.

In the $s_{i}\in\left\{ 0,1\right\} $ case, instead, we also generated
correlated patterns as combinations of sparse features $\phi^{\nu}$,
with $P\left(\phi_{i}^{\nu}\right)=f\,\delta\left(\phi_{i}^{\nu}-1\right)+\left(1-f\right)\delta\left(\phi_{i}^{\nu}\right)$,
chosen from a finite length dictionary $\mathcal{D}=\left\{ \phi^{\nu}\right\} _{\nu=1}^{L}$.
Every pattern contains a fixed number of features, $F$, and its components
can be written as: $\xi_{i}^{\mu}=\Theta\left(\sum_{\nu}c_{\nu}^{\mu}\phi_{i}^{\nu}\right)$,
with $c_{\nu}^{\mu}\in\left\{ 0,1\right\} $ determining whether the
feature $\nu$ appears in pattern $\mu$, and $\Theta\left(\cdot\right)$
is the Heaviside theta function, $\Theta\left(x\right)=1$ if $x>0$
and $\Theta\left(x\right)=0$ otherwise.

\section{TAP approximation in asymmetric sparse models\label{sec:TAP-approximation}}

In the heuristic version of DCM, the time-delayed correlations of
a network subject to varying external field intensities are needed
in order to update the model parameters. In our approach, we employed
a Monte Carlo scheme -- which relies solely on the network dynamics
-- for their evaluation, as a means to fulfill some basic biological
constraints. A better approximation though can be achieved with the
so-called TAP approach, consisting in a second order expansion around
a mean field limit, which can provide an estimation for the marginal
probabilities of the neuronal state variables. The related magnetizations
can then be used to compute approximate values for the pairwise correlations.

In what follows, we will apply the same procedure proposed in ref.\textbf{~}\citep{KappenMeanField}\textbf{
}for the $s_{i}\in\left\{ -1,+1\right\} $ case and the sequential
Glauber dynamics, to the $s_{i}\in\left\{ 0,1\right\} $ and the synchronized
dynamics case. Since we are dealing with an asymmetric model, where
the form of the joint probability distribution $P\left(s|\theta,J\right)$
is unknown, we have to assume a weakly interacting regime, with small
$\mathcal{O}\left(1/\sqrt{N}\right)$ couplings $J$, and in addition
to be close to a mean field model with a factorized distribution:
\begin{equation}
P^{\mathrm{MF}}\left(s|\theta^{\mathrm{MF}}\right)=\prod_{a=1}^{N}\frac{\exp\left(\theta_{a}^{\mathrm{MF}}s_{a}\right)}{1+\exp\left(\theta_{a}^{\mathrm{MF}}\right)}.
\end{equation}
We introduce the parametrization $\theta_{a}^{\mathrm{MF}}=\theta_{a}-d\theta_{a}$
where $d\theta_{a}$ is small and $\theta^{\mathrm{MF}}$ are the
parameters of the mean field model, which can be found by minimizing
the KL divergence:

\begin{equation}
KL\left[P||P^{\mathrm{MF}}\right]=\sum_{s}P\left(s|\theta,J\right)\log\left(\frac{P\left(s|\theta,J\right)}{P^{\mathrm{MF}}\left(s|\theta^{\mathrm{MF}}\right)}\right).
\end{equation}

The TAP approximation is obtained by performing a Taylor expansion
of the magnetizations $m_{a}=\sum_{s}P\left(s_{a}\right)s_{a}$ in
the small parameters $J_{jk}$ and $d\theta_{i}$ and applying the
matching condition $m_{a}-m_{a}^{\mathrm{MF}}=0$ for all $a\in\left\{ 1,\dots,N\right\} $
up to second order:

\begin{eqnarray}
0=m_{a}-m_{a}^{\mathrm{MF}} & \approx & \sum_{i}\left.\frac{\partial m_{a}}{\partial\theta_{i}}\right|_{\mathrm{MF}}d\theta_{i}+\sum_{i<j}\left.\frac{\partial m_{a}}{\partial J_{ij}}\right|_{\mathrm{MF}}dJ_{ij}+\\
 &  & +\sum_{ij}\left.\frac{\partial^{2}m_{a}}{\partial\theta_{i}\partial\theta_{j}}\right|_{\mathrm{MF}}d\theta_{i}d\theta_{j}+\sum_{i<j}\sum_{k<l}\left.\frac{\partial^{2}m_{a}}{\partial J_{ij}\partial J_{kl}}\right|_{\mathrm{MF}}dJ_{ij}dJ_{kl}+\nonumber \\
 &  & +2\sum_{i<j}\sum_{k}\left.\frac{\partial^{2}m_{a}}{\partial J_{ij}\partial\theta_{k}}\right|_{\mathrm{MF}}dJ_{ij}d\theta_{k}\nonumber 
\end{eqnarray}
After some calculations, the following derivatives, evaluated in correspondence
of the mean field probability distribution, are obtained:

\begin{eqnarray}
\left.\frac{\partial m_{a}}{\partial\theta_{i}}\right|_{\mathrm{MF}} & = & m_{a}\left(1-m_{a}\right)\delta_{ai}\\
\left.\frac{\partial m_{a}}{\partial J_{ij}}\right|_{\mathrm{MF}} & = & m_{j}m_{a}\left(1-m_{a}\right)\delta_{ai}\\
\left.\frac{\partial^{2}m_{a}}{\partial\theta_{i}\partial\theta_{j}}\right|_{\mathrm{MF}} & = & \left(m_{a}\left(1-m_{a}\right)^{2}-\left(m_{a}\right)^{2}\left(1-m_{a}\right)\right)\delta_{ai}\delta_{aj}\\
\left.\frac{\partial^{2}m_{a}}{\partial J_{ij}\partial\theta_{k}}\right|_{\mathrm{MF}} & = & m_{j}m_{a}\left(1-m_{j}\right)\left(1-m_{a}\right)\delta_{ai}\delta_{jk}+\\
 &  & +m_{j}\left[m_{a}\left(1-m_{a}\right)^{2}-\left(m_{a}\right)^{2}\left(1-m_{a}\right)\right]\delta_{ai}\delta_{ak}\nonumber \\
\left.\frac{\partial^{2}m_{a}}{\partial J_{ij}\partial J_{kl}}\right|_{\mathrm{MF}} & = & m_{j}m_{l}\left(1-m_{l}\right)m_{a}\left(1-m_{a}\right)\delta_{li}\delta_{ak}+\\
 &  & +m_{l}m_{j}\left(1-m_{j}\right)m_{a}\left(1-m_{a}\right)\delta_{jk}\delta_{ai}+\nonumber \\
 &  & +\left\langle s_{j}s_{l}\right\rangle _{\mathrm{MF}}\left(m_{a}\left(1-m_{a}\right)^{2}-\left(m_{a}\right)^{2}\left(1-m_{a}\right)\right)\delta_{ai}\delta_{ak}.\nonumber 
\end{eqnarray}

Using the identity $\left<s_{j}s_{l}\right>_{\mathrm{MF}}=\delta_{jl}m_{j}+\left(1-\delta_{jl}\right)m_{j}m_{l}$
and neglecting higher orders up to $\mathcal{O}\left(d\Theta^{2}\right)$,
the moment matching condition reads: 

\begin{eqnarray}
\theta_{a}^{\mathrm{MF}} & = & \theta_{a}+\sum_{j}m_{j}J_{aj}+\frac{1}{2}\left(1-2m_{a}\right)\sum_{j}\left(m_{j}\left(1-m_{j}\right)\right)J_{aj}^{2}.
\end{eqnarray}
This leads to the TAP equations for the single neuron marginal probabilities
when the sigmoid activation function is applied. The fixed point of
these equations can be found by recursion, with a \emph{crucial caveat},
namely that during the iterative procedure the time indices of the
magnetization appearing in the Onsager reaction term have to be chosen
carefully, according to:
\begin{equation}
m_{i}^{t+1}=\mbox{sigm}\left(\theta_{i}+\sum_{j}m_{j}^{t}J_{ij}-\left(m_{i}^{t-1}-\frac{1}{2}\right)\sum_{j}\left(m_{j}^{t}\left(1-m_{j}^{t}\right)\right)J_{ij}^{2}\right)
\end{equation}
where $\mathrm{sigm}\left(x\right)=\left(1+e^{-x}\right)^{-1}$ is
the sigmoid function. Note that in the model presented in the main
text the constant field $\theta_{i}$ is further decomposed into the
effect of an external field and of a negative threshold $\theta_{i}\to\lambda^{\mathrm{ext}}\left(\xi_{i}-\frac{1}{2}\right)-\tilde{\theta}_{i}$.

Once the magnetizations are estimated, one can calculate the time-delayed
correlations in the same TAP approximation. The dependence of these
correlations on the magnetizations can be derived starting from:

\begin{equation}
\left<s_{i}^{\prime}s_{j}\right>=\sum_{s}P\left(s\right)s_{j}\sum_{s_{i}^{\prime}}P\left(s_{i}^{\prime}|s\right)s_{i}^{\prime}.
\end{equation}
After expanding the sum over $s_{i}^{\prime}$, one simply obtains:
$\left<s_{i}^{\prime}s_{j}\right>=\left<\mbox{sigm}\left(h_{i}\right)s_{j}\right>$.
In order to simplify some of the following derivations, we first consider
the Taylor expansion of the connected time-delayed correlations: 
\begin{eqnarray}
\chi_{ij}^{D} & = & \left<s_{i}^{\prime}s_{j}\right>-m_{i}m_{j}=\left<s_{i}\left(\mbox{sigm}\left(h_{j}\right)-m_{j}\right)\right>.
\end{eqnarray}
In order to find an expression up to second order in $d\Theta$, we
need the following derivatives:

\begin{eqnarray}
\left.\frac{\partial\chi_{ba}^{D}}{\partial\theta_{i}}\right|_{\mathrm{MF}} & = & 0\\
\left.\frac{\partial\chi_{ba}^{D}}{\partial J_{ij}}\right|_{\mathrm{MF}} & = & m_{b}m_{a}\left(1-m_{b}\right)\left(1-m_{j}\right)\delta_{aj}\delta_{bi}\\
\left.\frac{\partial^{2}\chi_{ba}^{D}}{\partial\theta_{i}\partial\theta_{j}}\right|_{\mathrm{MF}} & = & 0\\
\left.\frac{\partial^{2}\chi_{ba}^{D}}{\partial J_{ij}\partial\theta_{k}}\right|_{\mathrm{MF}} & = & m_{a}m_{b}\left(1-m_{a}\right)\left(1-m_{b}\right)\left(1-2m_{b}\right)\delta_{aj}\delta_{bk}\delta_{bi}\\
\left.\frac{\partial^{2}\chi_{ba}^{D}}{\partial J_{ij}\partial J_{kl}}\right|_{\mathrm{MF}} & = & m_{a}m_{b}\left(1-m_{b}\right)\left(1-2m_{b}\right)\delta_{bk}\delta_{bi}\times\\
 &  & \times\left(\delta_{aj}\delta_{al}+\left(1-\delta_{aj}\right)\delta_{al}m_{j}\left(1-m_{a}\right)\right).
\end{eqnarray}

Using the following relation:
\begin{eqnarray}
\left\langle s_{a}s_{j}s_{l}\right\rangle _{\mathrm{MF}} & = & \delta_{aj}\left(\delta_{al}m_{a}+\left(1-\delta_{al}\right)m_{a}m_{l}\right)+\\
 &  & +\left(1-\delta_{aj}\right)\left(\delta_{al}m_{a}m_{j}+\left(1-\delta_{al}\right)m_{a}\left\langle s_{j}s_{l}\right\rangle _{\mathrm{MF}}\right),\nonumber 
\end{eqnarray}
we obtain the expression for the Taylor expansion up to second order:

\begin{equation}
\chi_{ij}^{D}=\left(m_{i}\left(1-m_{i}\right)\right)\left(m_{j}\left(1-m_{j}\right)\right)\left(J_{ij}+\frac{1}{2}\left(2m_{i}-1\right)\left(2m_{j}-1\right)\left(J_{ij}\right)^{2}\right),
\end{equation}
and therefore the final expression for the time-delayed correlations
reads:

\begin{eqnarray}
\left<s_{i}^{\prime}s_{j}\right> & = & \left(m_{i}\left(1-m_{i}\right)\right)\left(m_{j}\left(1-m_{j}\right)\right)\times\\
 &  & \times\left(J_{ij}+\frac{1}{2}\left(2m_{i}-1\right)\left(2m_{j}-1\right)\left(J_{ij}\right)^{2}\right)+m_{i}m_{j}.\nonumber 
\end{eqnarray}

\section{Visible-to-hidden directed synapses\label{sec:Visible-to-hidden}}

In the case of an architecture restricted to visible-to-hidden directed
connections, the network can be seen as a bipartite graph. At any
given time the state of a neuron in one of the two subsets is conditionally
dependent only on the state of the complementary subset of neurons
at the previous time:

\begin{equation}
P\left(s_{i}^{\prime},i\in\mathcal{V}|s\right)=P\left(s_{i}^{\prime},i\in\mathcal{V}|s_{\mathcal{H}}\right)
\end{equation}

\begin{equation}
P\left(s_{i}^{\prime},i\in\mathcal{H}|s\right)=P\left(s_{i}^{\prime},i\in\mathcal{H}|s_{\mathcal{V}}\right).
\end{equation}
Because of this property the joint conditional probability $P\left(s_{\mathcal{H}}|s_{\text{\ensuremath{\mathcal{V}}}}\right)$
can be factorized, and the clamped probability distribution can be
written explicitly:

\begin{equation}
P_{\mathrm{clamp}}\left(s;\xi\right)=\prod_{i\in\mathcal{V}}\delta_{s_{i},\xi_{i}}\prod_{j\in\mathcal{H}}P\left(s_{j}|s_{\mathcal{V}}=\xi\right).
\end{equation}

The learning rule can be derived from the minimization of the KL divergence
between the conditional probabilities obtained when the external field
intensity is $\lambda^{ext}=\infty$ and $\lambda^{ext}=0$, averaged
over the clamped probability distribution. By differentiating the
KL with respect to a hidden to visible synaptic coupling $J_{ij}$
with $i\in\mathcal{V},\,j\in H$, we get the following update rule:
\begin{flushleft}
\begin{eqnarray}
\Delta J_{ij} & \propto & P\left(s_{j}|s_{\mathcal{V}}=\xi\right)\xi_{i}s_{j}-\sum_{s\in\mathcal{H}}\prod_{k\in\mathcal{H}}P\left(s_{k}|s_{\mathcal{V}}=\xi\right)P\left(s_{i}^{\prime}|s_{\mathcal{H}}\right)s_{i}^{\prime}s_{j}.
\end{eqnarray}
As in the case of fully visible networks, the same increment would
be obtained if an on-line optimization of the pseudo-likelihood of
the model was instead implemented, except that now its estimation
implies an average over all the possible hidden neuronal states:
\par\end{flushleft}

\begin{equation}
\mathcal{L}\left(\left\{ \xi^{\mu}\right\} |J_{ij},\theta;\beta\right)=\frac{1}{M}\sum_{\mu=1}^{M}\sum_{i\in\mathcal{V}}\log\left(\sum_{s_{j}\in\mathcal{H}}\prod_{k\in\mathcal{H}}P\left(s_{k}|s_{\mathcal{V}}=\xi\right)P\left(s_{i}^{\prime}=\xi_{i}^{\mu}|s_{\mathcal{H}};\lambda^{ext}=0\right)\right).
\end{equation}

\subsection{MNIST Simulations}

Instead of trying to construct an artificial stimulus ensemble, we
use the MNIST database benchmark \citep{lecun1998mnist}, which consists
of $7\cdot10^{4}$ $28\times28$ grayscale images representing hand-written
digits in $10$ classes from $0$ to $9$. Images are sparse, with
an average luminosity of $\overline{f_{\xi}}=0.13066$ and every component
ranging in the interval $\xi_{i}^{\mu}\in\left[0,1\right]$. It is
rather natural to consider each pattern as an array of probabilities
of finding the corresponding neurons in the active state: we therefore
consider a stochastic network of $s_{i}\in\left\{ 0,1\right\} $ neurons,
whose visible component will be successively subject to an external
field corresponding to each one of the images, as before multiplied
by a field intensity $\lambda^{\mathrm{ext}}$. We hold out the last
$10^{4}$ images as a test set for the generalization performance,
and employ the first $6\cdot10^{4}$ images to learn the statistics
of the data. 

We consider an architecture with $\left|\mathcal{V}\right|=784+10$
visible neurons, plus $\left|\mathcal{H}\right|=1000$ hidden neurons
to guarantee a high representational capacity. The additional $10$
visible neurons, one for each digit, can allow the network to learn
input-output correlations: these neurons received a supervised input
indicating the correct label of the image during the learning phase
\citep{larochelle2012learningRBM_output_supervised}, and were present
in all the simulations described in the following; however, they are
exploited only for the classification task, being unessential for
the usual generative tasks. 

In order to point out how the DCM is able to deal with all the biological
constraints we are considering in this work, we offer the direct comparison
between two different learning models. In the first numerical experiment,
which serves as a benchmark reference, the network was trained in
the infinite signal limit ($\lambda_{\text{max}}\sim50$ is sufficient)
corresponding to the pseudo-likelihood method, with unconstrained
synapses, no inhibitory mechanism and using the time-delayed correlations
obtained in the TAP approximation. In the second numerical experiment,
meant to test the DCM rule in a more biologically plausible, we studied
a purely excitatory network and implemented the soft ``winner-takes-all''
inhibitory scheme, fixing an average hidden activity of $f_{h}=0.2$.
The network was required to learn from finite external fields ($\lambda_{\text{max}}=3$,
$\Delta\lambda=3/2$ and $\lambda_{\text{min}}=0$) and to estimate
the correlations simply through its own dynamics (specifically, we
considered $T=15$), as described in sec.~\ref{alg:learning_algo-1}.
In both experiments the networks cycled $2$ times through the $6\cdot10^{4}$
training images of the MNIST dataset.

In the second experiment setting, a very high level of noise can become
extremely detrimental: with large hidden layers the network is often
prone to falling into a completely symmetric state, with very poor
performance. One would instead want to exploit the initial randomness
in the synaptic couplings as a tool for breaking this otherwise problematic
symmetry between the hidden neurons. This can be either achieved by
choosing a lower temperature $\beta=30$ (we choose this setting,
to be compared with $\beta=2$ in the first experiment) or by rescaling
the initial random configuration of the synaptic couplings. 

\subsection*{Receptive fields}

\begin{figure}
\centering{}\includegraphics[viewport=0bp 100bp 720bp 470bp,clip,width=0.9\textwidth]{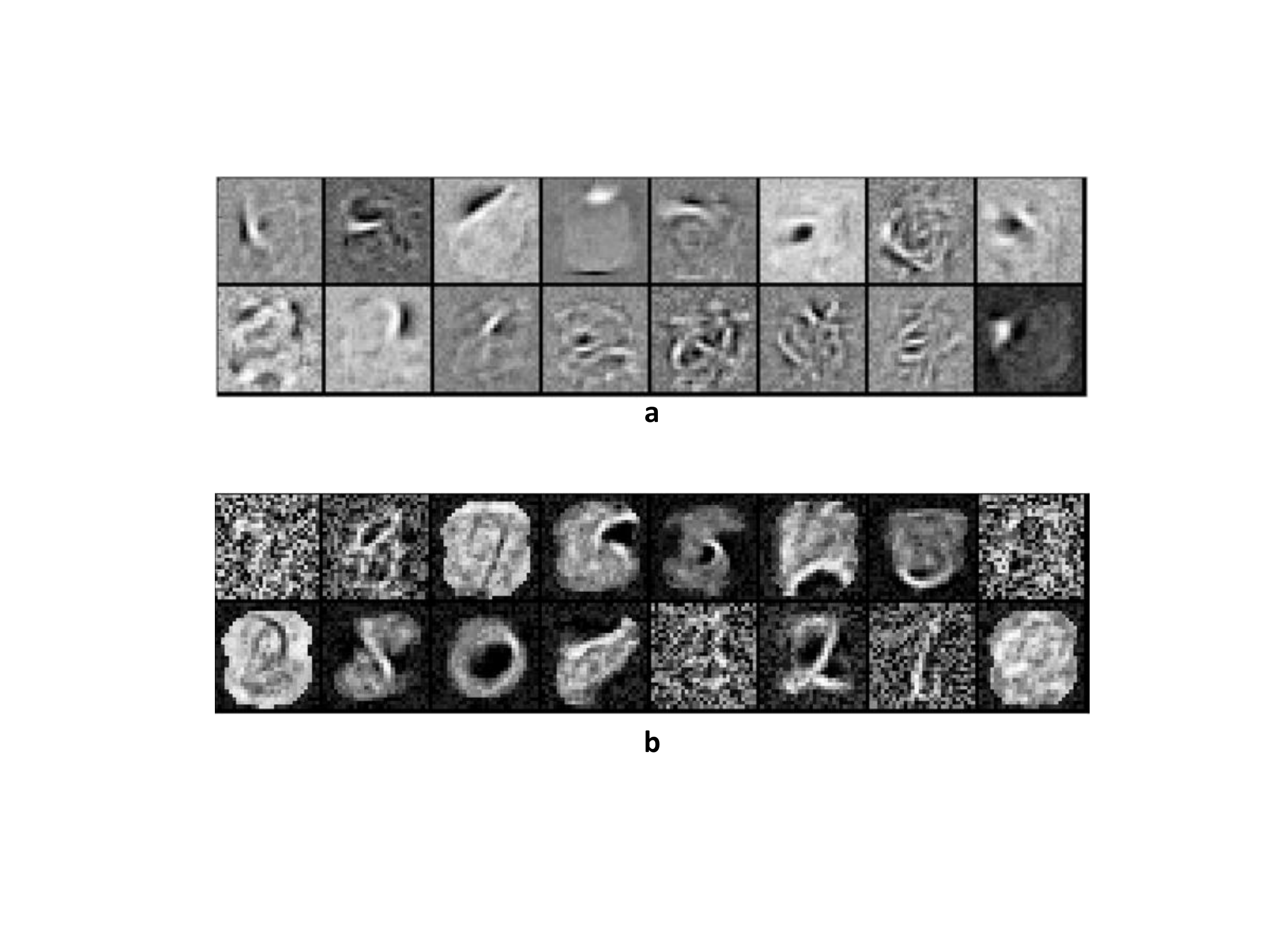}\caption{\label{fig:receptive_tap}\emph{Receptive fields of the hidden neurons.}
In this figure we show some of the receptive fields of the neurons
belonging to the hidden subset, in the two proposed experiments. \emph{a.}
Receptive fields learned in the first experiment, where the TAP approximation
was employed in the clamped limit with no inhibition. \emph{b.} Receptive
fields learned in the second experiments, where the correlation were
registered during the time evolution of the network, and where finite
time-dependent fields, constrained synapses and the ``winner-takes-all''
inhibition scheme where considered. }
\end{figure}

A first comparison of the learning performance in the two cases is
attained by visualizing the receptive fields of the hidden neurons,
which can show how each different hidden unit specializes in the detection
of a unique feature of the pattern set learned by the neural network.
The receptive fields of the hidden neurons are represented by the
synapses $J_{ij}$ with a fixed $i\in\mathcal{H}$ and $j$ running
through the visible indices $\mathcal{V}$. These arrays can be reorganized
as a $28\times28$ grayscale images as well, after renormalizing each
component in the interval $\left[0,1\right]$: the obtained image
represents, for any hidden unit, its optimal stimulus. A sample is
shown in fig.~\ref{fig:receptive_tap}.

It is apparent that most of the hidden units develop interesting internal
representations which can be interpreted as simple detectors for edges
of parts of single digits. Both experiments also show the presence
of a small fraction of extremely noisy features (that usually become
irrelevant since the threshold of the corresponding neuron raises
in order to inhibit its activation during the dynamics).

\subsection*{Generative tasks\label{subsec:Generative-tasks}}

\begin{figure}
\centering{}\includegraphics[viewport=0bp 150bp 720bp 460bp,clip,width=0.9\textwidth]{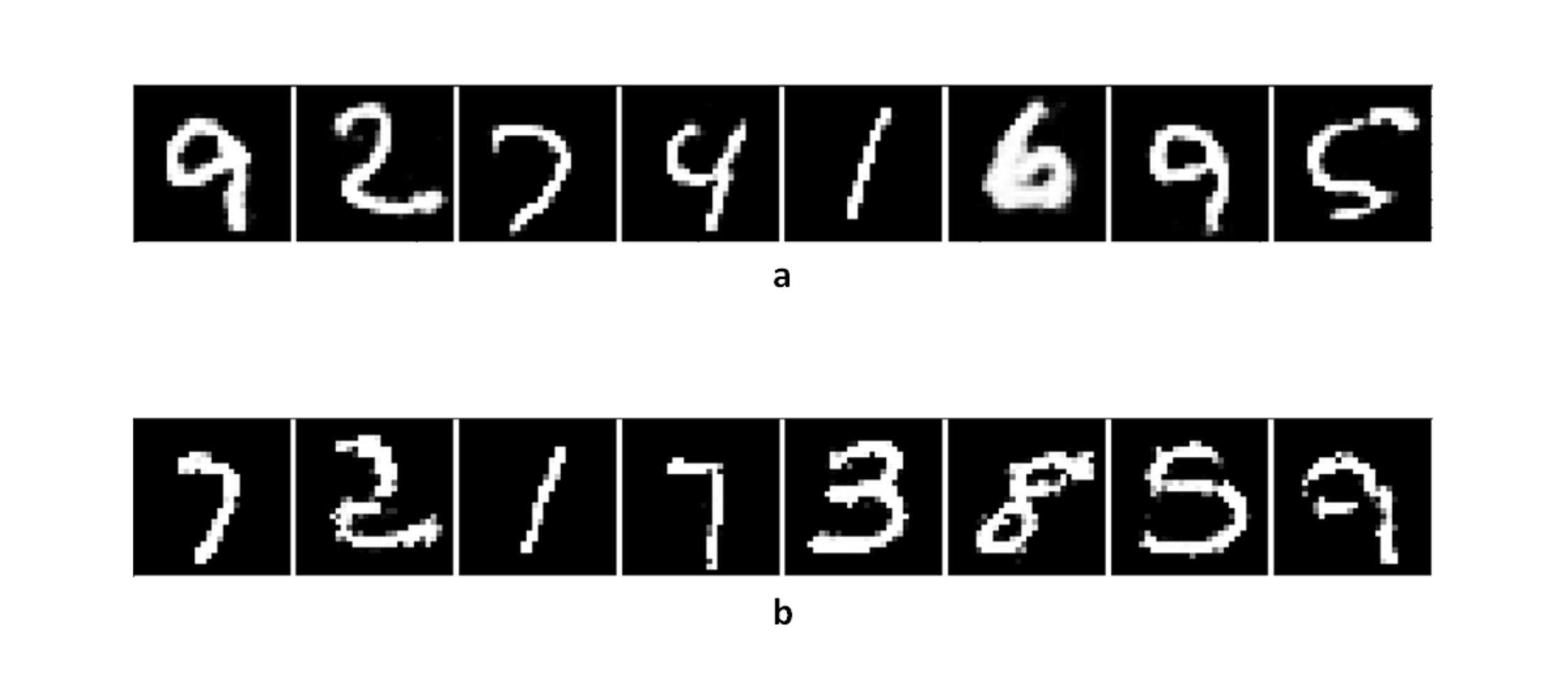}\caption{\label{fig:generation}\emph{Generation of samples. }The two series
of plots show the probability of obtaining an active state for the
visible neurons after $100$ time-steps of the dynamics, starting
from $8$ different initial states. \emph{a.} Samples generated in
the first experiment, where the TAP approximation was employed in
the clamped limit with no inhibition. \emph{b.} Samples generated
in the second experiments, where the correlation where registered
during the time evolution of the network, and where finite time dependent
fields, constrained synapses and the ``winner-takes-all'' inhibition
scheme where considered. The superior smoothness of the samples from
the first experiment is also due to the choice of a higher temperature
in the dynamics ($\beta_{1}=2$ against $\beta_{2}=30$, see Methods).}
\end{figure}
A better way of assessing the quality of the internal representation
of the learned dataset in the two experiments is to test the generative
properties of the networks. As shown in fig.~\ref{fig:generation},
we obtained some visible configurations from the steady-state distribution
of the models, generated according to the information learned from
the training images. The steady-state distribution is reached by the
dynamical evolution of the network when starting from a visible neuronal
state induced by one of the learned images. In order to initialize
the network, visible neurons are clamped with a very strong field
($\lambda^{\mathrm{ext}}=50$) in the direction the image and of the
correct label for an initial period of $30$ time steps. The field
on the first $784$ neurons is then removed, while the visible neurons
receiving the supervised stimulus are maintained clamped, and the
network is left evolving for some iterations. Keeping the output labels
clamped only mildly encourages the network to produce new samples
from the same category, and this small signal does not have a major
effect.

Alternatively the networks can be asked to generate the correct label
of a test image, presented to the network with a clamping signal.
In the first experiment, the output of the network was read directly
from the magnetizations obtained at convergence of the TAP equations
iterative procedure, by simply picking the maximum magnetization between
the ones corresponding to the visible neurons associated to the label
of each category. This network was able to reach a generalization
error rate of $2,76\%$. This result is far from state-of-the-art
classification performance (around $0.3\%$\textbf{ }\citep{lecun1998mnist}),
but is remarkably low if one takes into account the highly noisy environment
and the small size of the network. In the second experiment, the magnetizations
were instead explicitly registered during the dynamical evolution
of the network. In this case, the performance declined to a $7.74\%$
generalization error rate. This result is nevertheless of interest,
considering that the entire learning process was done without a clear
supervised signal and that the system was subject to a number of biological
requirements restraining the computational performance of the network.

%% file: main2.bbl
\begin{thebibliography}{10}

\bibitem{RollsDecoNoisy}
Gustavo Deco, Edmund~T Rolls, and Ranulfo Romo.
\newblock Stochastic dynamics as a principle of brain function.
\newblock {\em Progress in neurobiology}, 88(1):1--16, 2009.

\bibitem{cannon2010stochastic}
Robert~C Cannon, Cian O'Donnell, and Matthew~F Nolan.
\newblock Stochastic ion channel gating in dendritic neurons: morphology
  dependence and probabilistic synaptic activation of dendritic spikes.
\newblock {\em PLoS Comput Biol}, 6(8):e1000886, 2010.

\bibitem{flight2008synaptic}
Monica~Hoyos Flight.
\newblock Synaptic transmission: On the probability of release.
\newblock {\em Nature Reviews Neuroscience}, 9(10):736--737, 2008.

\bibitem{azouz1999cellular}
Rony Azouz and Charles~M Gray.
\newblock Cellular mechanisms contributing to response variability of cortical
  neurons in vivo.
\newblock {\em The Journal of neuroscience}, 19(6):2209--2223, 1999.

\bibitem{gerstner2002spiking}
Wulfram Gerstner and Werner~M Kistler.
\newblock {\em Spiking neuron models: Single neurons, populations, plasticity}.
\newblock Cambridge university press, 2002.

\bibitem{brascamp2006time}
Jan~W Brascamp, Raymond Van~Ee, Andre~J Noest, Richard~HAH Jacobs, and Albert~V
  van~den Berg.
\newblock The time course of binocular rivalry reveals a fundamental role of
  noise.
\newblock {\em Journal of vision}, 6(11):8--8, 2006.

\bibitem{buesing2011neural}
Lars Buesing, Johannes Bill, Bernhard Nessler, and Wolfgang Maass.
\newblock Neural dynamics as sampling: a model for stochastic computation in
  recurrent networks of spiking neurons.
\newblock {\em PLoS Comput Biol}, 7(11):e1002211, 2011.
\newblock URL: \url{http://dx.doi.org/10.1371/journal.pcbi.1002211}.

\bibitem{bliss1993synaptic}
Tim~VP Bliss, Graham~L Collingridge, et~al.
\newblock A synaptic model of memory: long-term potentiation in the
  hippocampus.
\newblock {\em Nature}, 361(6407):31--39, 1993.
\newblock URL:
  \url{http://smash.psych.nyu.edu/courses/spring16/learnmem/papers/Bliss1993.pdf}.

\bibitem{rogan1997fear}
Michael~T Rogan, Ursula~V St{\"a}ubli, and Joseph~E LeDoux.
\newblock Fear conditioning induces associative long-term potentiation in the
  amygdala.
\newblock {\em Nature}, 390(6660):604--607, 1997.
\newblock URL:
  \url{http://www.nature.com/nature/journal/v390/n6660/abs/390604a0.html}.

\bibitem{whitlock2006learning}
Jonathan~R Whitlock, Arnold~J Heynen, Marshall~G Shuler, and Mark~F Bear.
\newblock Learning induces long-term potentiation in the hippocampus.
\newblock {\em science}, 313(5790):1093--1097, 2006.
\newblock URL: \url{http://science.sciencemag.org/content/313/5790/1093}.

\bibitem{Ben-Yishai1995}
R.~Ben-Yishai, R.~L. Bar-Or, and H.~Sompolinsky.
\newblock Theory of orientation tuning in visual cortex.
\newblock {\em Proc Natl Acad Sci U S A}, 92(9):3844--3848, Apr 1995.
\newblock 7731993[pmid].
\newblock URL: \url{http://www.ncbi.nlm.nih.gov/pmc/articles/PMC42058/}.

\bibitem{ClopathEmergence}
Sadra Sadeh, Claudia Clopath, and Stefan Rotter.
\newblock Emergence of functional specificity in balanced networks with
  synaptic plasticity.
\newblock {\em PLoS Comput Biol}, 11(6):1--27, 06 2015.
\newblock URL: \url{http://dx.doi.org/10.1371%2Fjournal.pcbi.1004307}, \href
  {http://dx.doi.org/10.1371/journal.pcbi.1004307}
  {\path{doi:10.1371/journal.pcbi.1004307}}.

\bibitem{compte2000synaptic}
Albert Compte, Nicolas Brunel, Patricia~S Goldman-Rakic, and Xiao-Jing Wang.
\newblock Synaptic mechanisms and network dynamics underlying spatial working
  memory in a cortical network model.
\newblock {\em Cerebral Cortex}, 10(9):910--923, 2000.

\bibitem{zhang1996representation}
Kechen Zhang.
\newblock Representation of spatial orientation by the intrinsic dynamics of
  the head-direction cell ensemble: a theory.
\newblock {\em The journal of neuroscience}, 16(6):2112--2126, 1996.

\bibitem{Amit:1989:MBF:77051}
Daniel~J. Amit.
\newblock {\em Modeling Brain Function: the World of Attractor Neural
  Networks}.
\newblock Cambridge University Press, New York, NY, USA, 1989.

\bibitem{Hopfield1982}
J.~J. Hopfield.
\newblock Neural networks and physical systems with emergent collective
  computational abilities.
\newblock {\em Proc Natl Acad Sci U S A}, 79(8):2554--2558, Apr 1982.
\newblock 6953413[pmid].
\newblock URL: \url{http://www.ncbi.nlm.nih.gov/pmc/articles/PMC346238/}.

\bibitem{Hopfield1984}
J.~J. Hopfield.
\newblock Neurons with graded response have collective computational properties
  like those of two-state neurons.
\newblock {\em Proc Natl Acad Sci U S A}, 81(10):3088--3092, May 1984.
\newblock 6587342[pmid].
\newblock URL: \url{http://www.ncbi.nlm.nih.gov/pmc/articles/PMC345226/}.

\bibitem{amit1997model}
Daniel~J Amit and Nicolas Brunel.
\newblock Model of global spontaneous activity and local structured activity
  during delay periods in the cerebral cortex.
\newblock {\em Cerebral cortex}, 7(3):237--252, 1997.

\bibitem{hinton1983_BM}
Geoffrey~E Hinton and Terrence~J Sejnowski.
\newblock Optimal perceptual inference.
\newblock In {\em Proceedings of the IEEE conference on Computer Vision and
  Pattern Recognition}, pages 448--453. Citeseer, 1983.

\bibitem{hinton1985_BMlearning}
David~H Ackley, Geoffrey~E Hinton, and Terrence~J Sejnowski.
\newblock A learning algorithm for boltzmann machines.
\newblock {\em Cognitive science}, 9(1):147--169, 1985.
\newblock URL: \url{https://doi.org/10.1207/s15516709cog0901_7}.

\bibitem{carreira2005contrastive_RBM}
Miguel~A Carreira-Perpinan and Geoffrey~E Hinton.
\newblock On contrastive divergence learning.
\newblock In Robert~G. Cowell and Zoubin Ghahramani, editors, {\em Aistats},
  volume~10, pages 33--40. Society for Artificial Intelligence and Statistics,
  2005.

\bibitem{gabrie2015training_TAPBM}
Marylou Gabri{\'e}, Eric~W Tramel, and Florent Krzakala.
\newblock Training restricted boltzmann machine via the
  thouless-anderson-palmer free energy.
\newblock In Daniel~D. Lee, Masashi Sugiyama, Corinna Cortes, Neil Lawrence,
  and Roman Garnet, editors, {\em Advances in Neural Information Processing
  Systems}, pages 640--648. Neural Information Processing Systems Foundation,
  2015.

\bibitem{tanaka1998meanfield_BM}
Toshiyuki Tanaka.
\newblock Mean-field theory of boltzmann machine learning.
\newblock {\em Physical Review E}, 58(2):2302, 1998.

\bibitem{yasuda2012learning_PSL_Completed}
Muneki Yasuda, Junya Tannai, and Kazuyuki Tanaka.
\newblock Learning algorithm for boltzmann machines using max-product algorithm
  and pseudo-likelihood.
\newblock {\em Interdisciplinary information sciences}, 18(1):55--63, 2012.

\bibitem{kappen1998efficient_MF_2}
Hilbert~J. Kappen and Francisco de~Borja Rodr{\'\i}guez.
\newblock Efficient learning in boltzmann machines using linear response
  theory.
\newblock {\em Neural Computation}, 10(5):1137--1156, 1998.

\bibitem{KappenMeanField}
H.~J. Kappen and J.~J. Spanjers.
\newblock Mean field theory for asymmetric neural networks.
\newblock {\em Phys. Rev. E}, 61:5658--5663, May 2000.
\newblock URL: \url{http://link.aps.org/doi/10.1103/PhysRevE.61.5658}, \href
  {http://dx.doi.org/10.1103/PhysRevE.61.5658}
  {\path{doi:10.1103/PhysRevE.61.5658}}.

\bibitem{RoudiHertzMeanField}
Yasser Roudi and John Hertz.
\newblock Mean field theory for nonequilibrium network reconstruction.
\newblock {\em Phys. Rev. Lett.}, 106:048702, Jan 2011.
\newblock URL: \url{http://link.aps.org/doi/10.1103/PhysRevLett.106.048702},
  \href {http://dx.doi.org/10.1103/PhysRevLett.106.048702}
  {\path{doi:10.1103/PhysRevLett.106.048702}}.

\bibitem{mezard2011exact}
M~M{\'e}zard and J~Sakellariou.
\newblock Exact mean-field inference in asymmetric kinetic ising systems.
\newblock {\em Journal of Statistical Mechanics: Theory and Experiment},
  2011(07):L07001, 2011.

\bibitem{strata1999dale}
Piergiorgio Strata and Robin Harvey.
\newblock Dale's principle.
\newblock {\em Brain research bulletin}, 50(5):349--350, 1999.

\bibitem{catsigeras2013dale}
Eleonora Catsigeras.
\newblock Dale's principle is necessary for an optimal neuronal network's
  dynamics.
\newblock {\em arXiv preprint arXiv:1307.0597}, 2013.

\bibitem{baldassi2018inverse}
Carlo Baldassi, Federica Gerace, Luca Saglietti, and Riccardo Zecchina.
\newblock From inverse problems to learning: a statistical mechanics approach.
\newblock In {\em Journal of Physics: Conference Series}, volume 955, page
  012001. IOP Publishing, 2018.

\bibitem{nguyen2017inverse}
H~Chau Nguyen, Riccardo Zecchina, and Johannes Berg.
\newblock Inverse statistical problems: from the inverse ising problem to data
  science.
\newblock {\em Advances in Physics}, 66(3):197--261, 2017.
\newblock URL: \url{https://doi.org/10.1080/00018732.2017.1341604}.

\bibitem{aurell2012inverse_PSL}
Erik Aurell and Magnus Ekeberg.
\newblock Inverse ising inference using all the data.
\newblock {\em Physical review letters}, 108(9):090201, 2012.

\bibitem{alemithreshold}
Alireza Alemi, Carlo Baldassi, Nicolas Brunel, and Riccardo Zecchina.
\newblock A three-threshold learning rule approaches the maximal capacity of
  recurrent neural networks.
\newblock {\em PLOS Computational Biology}, 11(8):1--23, 08 2015.
\newblock URL: \url{http://dx.doi.org/10.1371%2Fjournal.pcbi.1004439}, \href
  {http://dx.doi.org/10.1371/journal.pcbi.1004439}
  {\path{doi:10.1371/journal.pcbi.1004439}}.

\bibitem{binas2014learning}
Jonathan Binas, Ueli Rutishauser, Giacomo Indiveri, and Michael Pfeiffer.
\newblock Learning and stabilization of winner-take-all dynamics through
  interacting excitatory and inhibitory plasticity.
\newblock {\em Frontiers in computational neuroscience}, 8:68, 2014.
\newblock URL:
  \url{http://journal.frontiersin.org/article/10.3389/fncom.2014.00068/full}.

\bibitem{douglas1989canonical}
Rodney~J Douglas, Kevan~AC Martin, and David Whitteridge.
\newblock A canonical microcircuit for neocortex.
\newblock {\em Neural computation}, 1(4):480--488, 1989.
\newblock URL:
  \url{http://www.mitpressjournals.org/doi/abs/10.1162/neco.1989.1.4.480#.WCyY6bUc10w},
  \href {http://dx.doi.org/10.1162/neco.1989.1.4.480}
  {\path{doi:10.1162/neco.1989.1.4.480}}.

\bibitem{mountcastle1997columnar}
Vernon~B Mountcastle.
\newblock The columnar organization of the neocortex.
\newblock {\em Brain}, 120(4):701--722, 1997.
\newblock URL: \url{http://dx.doi.org/10.1093/brain/120.4.701}.

\bibitem{binzegger2004quantitative}
Tom Binzegger, Rodney~J Douglas, and Kevan~AC Martin.
\newblock A quantitative map of the circuit of cat primary visual cortex.
\newblock {\em The Journal of Neuroscience}, 24(39):8441--8453, 2004.
\newblock URL: \url{http://dx.doi.org/10.1523/JNEUROSCI.1400-04.2004}.

\bibitem{douglas2007recurrent}
Rodney~J Douglas and Kevan~AC Martin.
\newblock Recurrent neuronal circuits in the neocortex.
\newblock {\em Current Biology}, 17(13):R496--R500, 2007.
\newblock URL: \url{http://dx.doi.org/10.1016/j.cub.2007.04.024}.

\bibitem{carandini2012normalization}
Matteo Carandini and David~J Heeger.
\newblock Normalization as a canonical neural computation.
\newblock {\em Nature Reviews Neuroscience}, 13(1):51--62, 2012.
\newblock URL: \url{http://www.nature.com/nrn/journal/v13/n1/abs/nrn3136.html}.

\bibitem{handrich2009biologically}
Sebastian Handrich, Andreas Herzog, Andreas Wolf, and Christoph~S Herrmann.
\newblock A biologically plausible winner-takes-all architecture.
\newblock In {\em International Conference on Intelligent Computing}, pages
  315--326. Springer, 2009.
\newblock URL:
  \url{http://link.springer.com/chapter/10.1007/978-3-642-04020-7_34}, \href
  {http://dx.doi.org/10.1007/978-3-642-04020-7_34}
  {\path{doi:10.1007/978-3-642-04020-7_34}}.

\bibitem{mao2007dynamics}
Zhi-Hong Mao and Steve~G Massaquoi.
\newblock Dynamics of winner-take-all competition in recurrent neural networks
  with lateral inhibition.
\newblock {\em IEEE transactions on neural networks}, 18(1):55--69, 2007.
\newblock URL:
  \url{http://ieeexplore.ieee.org/document/4049830/?arnumber=4049830}.

\bibitem{lynch2016computational}
Nancy Lynch, Cameron Musco, and Merav Parter.
\newblock Computational tradeoffs in biological neural networks:
  Self-stabilizing winner-take-all networks.
\newblock {\em arXiv preprint arXiv:1610.02084}, 2016.
\newblock URL: \url{https://arxiv.org/abs/1610.02084}.

\bibitem{oster2006spiking}
Matthias Oster and Shih-Chii Liu.
\newblock Spiking inputs to a winner-take-all network.
\newblock {\em Advances in Neural Information Processing Systems}, 18:1051,
  2006.

\bibitem{fang1996dynamics}
Yuguang Fang, Michael~A Cohen, and Thomas~G Kincaid.
\newblock Dynamics of a winner-take-all neural network.
\newblock {\em Neural Networks}, 9(7):1141--1154, 1996.
\newblock URL: \url{http://dx.doi.org/10.1016/0893-6080(96)00019-6}.

\bibitem{amit1987information_Hebb_biased}
Daniel~J Amit, Hanoch Gutfreund, and Haim Sompolinsky.
\newblock Information storage in neural networks with low levels of activity.
\newblock {\em Physical Review A}, 35(5):2293, 1987.

\bibitem{nadal1986networks}
JP~Nadal, G~Toulouse, JP~Changeux, and S~Dehaene.
\newblock Networks of formal neurons and memory palimpsests.
\newblock {\em EPL (Europhysics Letters)}, 1(10):535, 1986.
\newblock URL: \url{https://doi.org/10.1209/0295-5075/1/10/008}.

\bibitem{storkey1998palimpsest}
Amos~J Storkey and Romain Valabregue.
\newblock The basins of attraction of a new hopfield learning rule.
\newblock {\em Neural Networks}, 12(6):869--876, 1999.

\bibitem{parisi1986memory}
Giorgio Parisi.
\newblock A memory which forgets.
\newblock {\em Journal of Physics A: Mathematical and General}, 19(10):L617,
  1986.

\bibitem{salakhutdinov2008learning_Hardness_BM}
Ruslan Salakhutdinov.
\newblock Learning and evaluating boltzmann machines.
\newblock {\em Tech. Rep., Technical Report UTML TR 2008-002, Department of
  Computer Science, University of Toronto}, 2008.

\bibitem{hinton2006_RBM}
Geoffrey~E Hinton and Ruslan~R Salakhutdinov.
\newblock Reducing the dimensionality of data with neural networks.
\newblock {\em science}, 313(5786):504--507, 2006.

\bibitem{thouless1977solution_TAP}
David~J Thouless, Philip~W Anderson, and Robert~G Palmer.
\newblock Solution of'solvable model of a spin glass'.
\newblock {\em Philosophical Magazine}, 35(3):593--601, 1977.

\bibitem{lecun1998mnist}
Yann LeCun.
\newblock {The MNIST database of handwritten digits}.
\newblock \url{http://yann.lecun.com/exdb/mnist/}.

\bibitem{Reinhart2011}
R.~Felix Reinhart and Jochen~J. Steil.
\newblock A constrained regularization approach for input-driven recurrent
  neural networks.
\newblock {\em Differential Equations and Dynamical Systems}, 19(1):27--46,
  2011.
\newblock URL: \url{http://dx.doi.org/10.1007/s12591-010-0067-x}, \href
  {http://dx.doi.org/10.1007/s12591-010-0067-x}
  {\path{doi:10.1007/s12591-010-0067-x}}.

\bibitem{Mayer2004}
Norbert~M. Mayer and Matthew Browne.
\newblock {\em Echo State Networks and Self-Prediction}, pages 40--48.
\newblock Springer Berlin Heidelberg, Berlin, Heidelberg, 2004.
\newblock URL: \url{http://dx.doi.org/10.1007/978-3-540-27835-1_4}, \href
  {http://dx.doi.org/10.1007/978-3-540-27835-1_4}
  {\path{doi:10.1007/978-3-540-27835-1_4}}.

\bibitem{sussilloTransferring}
David Sussillo and L.F. Abbott.
\newblock Transferring learning from external to internal weights in echo-state
  networks with sparse connectivity.
\newblock {\em PLoS ONE}, 7(5):1--8, 05 2012.
\newblock URL: \url{http://dx.doi.org/10.1371%2Fjournal.pone.0037372}, \href
  {http://dx.doi.org/10.1371/journal.pone.0037372}
  {\path{doi:10.1371/journal.pone.0037372}}.

\bibitem{jaeger2014controlling}
Herbert Jaeger.
\newblock Controlling recurrent neural networks by conceptors.
\newblock {\em arXiv preprint arXiv:1403.3369}, 2014.

\bibitem{abbott2016building}
LF~Abbott, Brian DePasquale, and Raoul-Martin Memmesheimer.
\newblock Building functional networks of spiking model neurons.
\newblock {\em Nature neuroscience}, 19(3):350--355, 2016.

\bibitem{depasquale2016using}
Brian DePasquale, Mark~M Churchland, and LF~Abbott.
\newblock Using firing-rate dynamics to train recurrent networks of spiking
  model neurons.
\newblock {\em arXiv preprint arXiv:1601.07620}, 2016.

\bibitem{scellier2017equilibrium}
Benjamin Scellier and Yoshua Bengio.
\newblock Equilibrium propagation: Bridging the gap between energy-based models
  and backpropagation.
\newblock {\em Frontiers in computational neuroscience}, 11:24, 2017.
\newblock URL: \url{https://doi.org/10.3389/fncom.2017.00024}.

\bibitem{braunstein2011inference}
Alfredo Braunstein, Abolfazl Ramezanpour, Riccardo Zecchina, and Pan Zhang.
\newblock Inference and learning in sparse systems with multiple states.
\newblock {\em Physical Review E}, 83(5):056114, 2011.
\newblock URL: \url{https://doi.org/10.1103/PhysRevE.83.056114}.

\bibitem{gidas1988consistency_ML_PML}
Basilis Gidas.
\newblock Consistency of maximum likelihood and pseudo-likelihood estimators
  for gibbs distributions.
\newblock In {\em Stochastic differential systems, stochastic control theory
  and applications}, pages 129--145. Springer, 1988.

\bibitem{gardnerSpace}
E~Gardner.
\newblock The space of interactions in neural network models.
\newblock {\em Journal of Physics A: Mathematical and General}, 21(1):257,
  1988.
\newblock URL: \url{http://stacks.iop.org/0305-4470/21/i=1/a=030}.

\bibitem{engel2001statistical}
Andreas Engel.
\newblock {\em Statistical mechanics of learning}.
\newblock Cambridge University Press, 2001.

\bibitem{fontaine2014spike}
Bertrand Fontaine, Jos{\'e}~Luis Pe{\~n}a, and Romain Brette.
\newblock Spike-threshold adaptation predicted by membrane potential dynamics
  in vivo.
\newblock {\em PLoS Comput Biol}, 10(4):e1003560, 2014.
\newblock URL:
  \url{http://journals.plos.org/ploscompbiol/article?id=10.1371/journal.pcbi.1003560}.

\bibitem{huang2016adaptive}
Chao Huang, Andrey Resnik, Tansu Celikel, and Bernhard Englitz.
\newblock Adaptive spike threshold enables robust and temporally precise
  neuronal encoding.
\newblock {\em PLoS Comput Biol}, 12(6):e1004984, 2016.
\newblock URL:
  \url{http://journals.plos.org/ploscompbiol/article?id=10.1371/journal.pcbi.1004984}.

\bibitem{amit1987statistical}
Daniel~J Amit, Hanoch Gutfreund, and Haim Sompolinsky.
\newblock Statistical mechanics of neural networks near saturation.
\newblock {\em Annals of physics}, 173(1):30--67, 1987.
\newblock URL:
  \url{http://www.sciencedirect.com/science/article/pii/0003491687900923},
  \href {http://dx.doi.org/doi:10.1016/0003-4916(87)90092-3}
  {\path{doi:doi:10.1016/0003-4916(87)90092-3}}.

\bibitem{larochelle2012learningRBM_output_supervised}
Hugo Larochelle, Michael Mandel, Razvan Pascanu, and Yoshua Bengio.
\newblock Learning algorithms for the classification restricted boltzmann
  machine.
\newblock {\em Journal of Machine Learning Research}, 13(Mar):643--669, 2012.

\end{thebibliography}
